\documentclass[12pt,a4paper,english,onecolumn,draftcls]{IEEEtran}
\usepackage[T1]{fontenc}
\usepackage[latin9]{inputenc}
\usepackage{xcolor}
\usepackage{babel}
\usepackage{amsmath}
\usepackage{amssymb}
\usepackage{stackrel}
\usepackage{graphicx}
\usepackage{setspace}
\PassOptionsToPackage{normalem}{ulem}
\usepackage{ulem}
\makeatletter

\providecolor{lyxadded}{rgb}{0,0,1}
\providecolor{lyxdeleted}{rgb}{1,0,0}
\DeclareRobustCommand{\lyxadded}[3]{{\texorpdfstring{\color{lyxadded}{}}{}#3}}

\DeclareRobustCommand{\lyxsout}[1]{\ifx\\#1\else\sout{#1}\fi}

\usepackage{subfigure}
\usepackage{epstopdf}
\usepackage{cite}
\usepackage{balance}

\makeatother

\begin{document}

\title{On the Energy-Efficient Deployment for Ultra-Dense Heterogeneous
Networks with NLoS and LoS Transmissions}

\author{\noindent {\normalsize{}Bin Yang, }\textit{\normalsize{}Student Member,
IEEE}{\normalsize{}, Guoqiang Mao, }\lyxadded{Bean}{Thu Nov 23 02:37:13 2017}{\emph{\normalsize{}Fellow}}\emph{\normalsize{},
IEEE, }{\normalsize{}Xiaohu Ge, }\textit{\normalsize{}Senior Member,
IEEE,}{\normalsize{} Ming Ding, }\textit{\normalsize{}Senior Member,
IEEE, }{\normalsize{}Xuan Yang} \thanks{Correspondence author: Prof. Xiaohu Ge (Email: xhge@mail.hust.edu.cn,
Tel: +86-27-87557941 ext 822).}\thanks{Bin Yang, Xiaohu Ge and Xuan Yang are with the School of Electronic
Information and Communications, Huazhong University of Science and
Technology, China (Email: yangbin@hust.edu.cn, xhge@mail.hust.edu.cn,
xuan\_yan9@163.com). }\thanks{Guoqiang Mao is with the School of Computing and Communication, The
University of Technology Sydney, Australia (e-mail: g.mao@ieee.org). }\thanks{Ming Ding is with Data61, CSIRO, Australia (Email: ming.ding@data61.csiro.au). }}
\maketitle
\begin{abstract}
We investigate network performance of ultra-dense heterogeneous networks
(HetNets) and study the maximum energy-efficient base station (BS)
deployment incorporating probabilistic non-line-of-sight (NLoS) and
line-of-sight (LoS) transmissions. First, we develop an analytical
framework with the maximum instantaneous received power (MIRP) and
the maximum average received power (MARP) association schemes to model
the coverage probability and related performance metrics, \emph{e.g.},
the potential throughput (PT) and the energy efficiency (EE). Second,
we formulate two optimization problems to achieve the maximum energy-efficient
deployment solution with \lyxadded{Bean}{Thu Dec  7 09:34:18 2017}{specific}
service criteria. Simulation results show that there are tradeoffs
among the coverage probability, the total power consumption\lyxadded{Bean}{Thu Dec  7 09:34:35 2017}{,}
and the EE. To be specific, the maximum coverage probability with
ideal power consumption is superior to that with practical power consumption
when the total power constrain\lyxadded{Bean}{Thu Dec  7 05:50:34 2017}{t}
is small and inferior to that with practical power consumption when
the total power constrain\lyxadded{Bean}{Thu Dec  7 05:50:39 2017}{t}
becomes large. Moreover, the maximum EE is a decreasing function with
respect to the coverage probability constrain\lyxadded{Bean}{Thu Dec  7 05:50:44 2017}{t}.
\end{abstract}


\section{Introduction}

Ultra-dense deployment of small cell base stations (BSs), relay nodes,
and distributed antennas is considered as a de facto solution for
realizing the significant performance improvements needed to accommodate
the overwhelming future mobile traffic demand\lyxadded{Bean}{Thu Dec  7 12:47:33 2017}{
\cite{Ge165G}}. Traditional network expansion techniques like cell
splitting are often utilized by telecom operators to achieve the expected
throughput, which \lyxadded{Bean}{Thu Dec  7 09:36:05 2017}{is} less
efficient and proven not to keep up with the pace of traffic proliferation
in the near future. Heterogeneous networks (HetNets) then become a
promising and attractive network architecture to alleviate the problem.
``HetNets'' is a broad term that refers to the coexistence of different
networks (\emph{e.g.}, traditional macrocells and small cell networks
like femtocells and picocells), each of them constituting a network
tier. Due to differences in deployment, BSs in different tiers may
have different transmit powers, radio access technologies, fading
environments and spatial densities. HetNets are envisioned to change
the existing network architectures and have been introduced in the
LTE-Advanced standardization \cite{LopezPerez15Towards,Andrews11A}. 

Massive work has been done in HetNets scenario mainly related \lyxadded{Bean}{Thu Dec  7 09:37:48 2017}{to}
cell association scheme \cite{Liu16User,Bethanabhotla16Optimal,Wang16Joint},
cache-enabled networks \cite{Yang16Analysis}, physical layer security
\cite{Wang16Physical}, \emph{etc}. In \cite{Liu16User}, the pertinent
user association algorithms designed for HetNets, massive MIMO networks,
mmWave scenarios and energy harvesting networks have been surveyed
for the future fifth generation (5G) networks. Bethanabhotla\emph{
et al.} \cite{Bethanabhotla16Optimal} investigated the optimal user-cell
association problem for massive MIMO HetNets and illustrated how massive
MIMO \lyxadded{Bean}{Thu Dec  7 09:38:30 2017}{could} also provide
nontrivial advantages at the system level. The joint downlink cell
association and wireless backhaul bandwidth allocation in a two-tier
HetNet is studied in \cite{Wang16Joint}. In \cite{Yang16Analysis},
Yang \emph{et al.} aimed to model and evaluate the performance of
the wireless HetNet where the radio access network (RAN) caching and
device-to-device (D2D) caching coexist. The physical layer security
of HetNets where the locations of all BSs, \lyxadded{Bean}{Wed Dec  6 06:55:26 2017}{mobile
users (}MUs\lyxadded{Bean}{Wed Dec  6 06:55:28 2017}{)} and eavesdroppers
are modeled as independent homogeneous PPPs in \cite{Wang16Physical}.

From the mobile operators point of view, the commercial viability
of network \lyxadded{Bean}{Thu Dec  7 09:39:48 2017}{densification}
depends on the underlying capital and operational expenditure \cite{Humar11Rethinking}.
While the former cost may be covered by taking up a high volume of
customers, with the rapid rise in the price of energy, and given that
BSs are particularly power-hungry, energy efficiency (EE) has become
an increasingly crucial factor for the success of dense HetNets \cite{Shojaeifard16Stochastic}.
Recently, loads of work \cite{Niu17Energy,Shahid17Self,Wu17Energy,Yang17Energy,Zhangke16Energy}
has investigated the EE in the 5G network scenarios. In \cite{Niu17Energy},
Niu \emph{et al.} investigated the problem of minimizing the energy
consumption via optimizing concurrent transmission scheduling and
power control for the mmWave backhauling of small cells densely deployed
in HetNets. A self-organized cross-layer optimization for enhancing
the EE of the D2D communications without creating harmful impact on
other tiers by employing a non-cooperative game in a three-tier HetNet
is proposed in \cite{Shahid17Self}. To jointly optimize the EE and
video quality, Wu \emph{et al.} \cite{Wu17Energy} presented an energy-quality
aware bandwidth aggregation scheme. In \cite{Yang17Energy}, Yang
\emph{et al.} investigated the energy-efficient resource allocation
problem for downlink heterogeneous OFDMA networks. The mobile edge
computing offloading mechanisms are studied in 5G HetNets \cite{Zhangke16Energy}.

Different from most prior work analyzing network performance where
the propagation path loss between the BSs and the MUs\lyxadded{Bean}{Wed Dec  6 06:55:59 2017}{
}follows the same power-law model, in this paper we consider the co-existence
of both non-line-of-sight (NLoS) and line-of-sight (LoS) transmissions,
which frequently occur in \emph{urban areas}. More specifically, for
a randomly selected MU, BSs deployed according to a homogeneous Poisson
point process (PPP) are divided into two categories, \emph{i.e.},
NLoS BSs and LoS BSs, depending on the distance between BSs and MUs.
\lyxadded{Bean}{Wed Dec  6 03:22:28 2017}{It is well known that LoS
transmission may occur when the distance between a transmitter and
a receiver is small, and NLoS transmission is common in office environments
and central business districts. Moreover, as the trend of ultra-dense
network deployment, the distance between a transmitter and a receiver
decreases, the probability that a LoS path exists between them increases,
thereby causing a transition from NLoS transmission to LoS transmission
with a higher probability \cite{Ding16Performance}. }In this context,
Ding \emph{et al.} \cite{Ding16Performance} studied the coverage
and capacity performance by using a multi-slop path loss model incorporating
probabilistic NLoS and LoS transmissions. The coverage and capacity
performance in millimeter wave cellular networks are studied in \cite{DiRenzo15StochasticJ,Singh15Tractable,Bai15Coverage}.
In \cite{DiRenzo15StochasticJ}, a three-state statistical model for
each link was assumed, in which a link can either be in a\lyxadded{Bean}{Thu Dec  7 09:22:06 2017}{n}
NLoS, LoS or an outage state. In \cite{Singh15Tractable}, self-backhauled
millimeter wave cellular networks are characterized assuming a cell
association scheme based on the smallest path loss. However, both
\cite{DiRenzo15StochasticJ} and \cite{Singh15Tractable} assume a
noise-limited network, ignoring inter-cell interference, which may
not be very practical since modern wireless networks work in the interference-limited
region. \lyxadded{Bean}{Wed Dec  6 03:21:13 2017}{In \cite{Bai15Coverage},
the coverage probability and capacity were calculated in a millimeter
wave cellular network based on the smallest path loss cell association
model assuming multi-path fading modeled as Nakagami-$m$ fading,
respectively. However, shadowing was ignored in their models, which
may not be very practical for an ultra-dense heterogeneous network. }

In contrast to prior work, we investigate the HetNets in a more realistic
scenario, \emph{i.e.}, NLoS and LoS transmissions\lyxadded{Bean}{Wed Dec  6 06:44:38 2017}{
in desired signal and interference signal} are both considered. Besides,
we also explore the optimal BS deployment under the quality of service
(QoS) constrain\lyxadded{Bean}{Thu Dec  7 05:50:48 2017}{t}. The main
contributions of this paper are summarized as follows:
\begin{enumerate}
\item \lyxadded{Bean}{Wed Dec  6 06:45:56 2017}{\textbf{A }}\textbf{unified
framework:} We propose a unified framework, in which the user association
strategies based on the maximum instantaneous received power (MIRP)
and the maximum average received power (MARP) can be studied, assuming
log-normal shadowing, Rayleigh fading and incorporating probabilistic
NLoS and LoS transmissions.
\item \textbf{Performance optimization:} We formulate two optimization problems
under different QoS constrain\lyxadded{Bean}{Thu Dec  7 05:50:51 2017}{t}s,
\emph{i.e.}, the maximal total power consumption and the minimal coverage
probability. Utilizing solutions of the above optimization problems,
the maximum energy-efficient BS deployment is obtained.
\item \textbf{Network design insights:} We compare the optimal BS deployment
strategies in different network scenarios, \emph{i.e.}, assuming the
fixed transmit power, the density-dependent transmit power, with and
without considering the static power consumption in BSs. Through our
results, the maximum coverage probability with ideal power consumption
is superior to that with practical power consumption when the total
power constrain\lyxadded{Bean}{Thu Dec  7 05:50:54 2017}{t} is small
and inferior to that with practical power consumption when the total
power constrain\lyxadded{Bean}{Thu Dec  7 05:50:58 2017}{t} becomes
large. Moreover, the maximum EE is a decreasing function with respect
to the coverage probability constrain\lyxadded{Bean}{Thu Dec  7 05:51:01 2017}{t}.
\end{enumerate}
~~The \lyxadded{Bean}{Thu Dec  7 09:50:00 2017}{remainder} of this
paper is organized as follows. Section \ref{sec:System-Model} introduces
the system model, network assumptions\lyxadded{Bean}{Thu Dec  7 09:50:30 2017}{,}
and performance metrics. In section \ref{sec:Performance Analysis},
the coverage probability, the potential throughput (PT) and the EE
of the HetNets are derived with the MIRP and the MARP association
schemes, respectively. In Section \ref{sec:Performance Optimization},
two optimization problems for energy-efficient BS deployment are formulated.
In Section \ref{sec:Results-and-Insights}, the analytical results
are validated via Monte Carlo simulations. Besides, the insights of
BS deployment are studied. Finally, Section \ref{sec:Conclusions-and-Future}
concludes this paper and discusses possible future work.

\section{\label{sec:System-Model}System Model}

In this paper, a $K$-tier HetNet is considered, which consists of
macrocells, picocells, femtocells, \emph{etc.} BSs of each tier are
assumed to be spatially distributed on the infinite plane and locations
of BSs follow independent homogeneous Poisson point processes (HPPPs)
denoted by $\Phi_{k}=\left\{ \boldsymbol{X}_{k,i}\right\} $ with
\lyxadded{Bean}{Fri Nov 17 03:13:11 2017}{a} density (\emph{aka} intensity)
$\lambda_{k}$, $k\in\left\{ 1,2,\cdots,K\right\} \overset{\textrm{def}}{=}\mathcal{K}$
\footnote{$x\overset{\textrm{def}}{=}y$ means $x$ is defined to be another
name for $y$.}, where $\boldsymbol{X}_{k,i}$ denotes the location of BS in the
$k$-th tier. MUs are deployed according to another independent HPPP
denoted by $\Phi_{u}$ with \lyxadded{Bean}{Fri Nov 17 03:14:07 2017}{a}
density $\lambda_{u}$ ($\lambda_{u}\gg\lambda_{k}$). BSs belonging
to the same tier transmit using the same constant power $P_{k}$ and
sharing the same bandwidth. Besides, within a cell assume that each
MU uses orthogonal multiple access method to connect to a serving
BS for downlink and uplink transmissions and therefore there is no
intra-cell interference in the analysis of our paper. However, adjacent
BSs which are not serving the connected MU may cause inter-cell interference
which is the main focus of this paper. It is further assumed that
each MU can possibly associate with a BS belonging to any tier, \emph{i.e.},
\emph{open access }policy is employed.

\lyxadded{Bean}{Wed Dec  6 07:08:14 2017}{Without loss of generality
and from the Slivnyak's Theorem \cite{Haenggi12StochasticB}, we consider
the \emph{typical MU}which is usually assumed to be located at the
origin, as the focus of our performance analysis.}

\subsection{\label{subsec:Signal-Propagation-Model}Signal Propagation Model}

The long-distance signal attenuation in tier $k$ is modeled by a
monotone, non-increasing and continuous path loss function $l_{k}:\left[0,\infty\right]\mapsto\left[0,\infty\right]$
and $l_{k}$ decays to zero asymptotically. The fast fading coefficient
for the wireless link between a BS \textbf{$\boldsymbol{X}_{k,i}\in\Phi_{k}$}
and the typical MU is denoted as $h_{\boldsymbol{X}_{k,i}}$. $\left\{ h_{\boldsymbol{X}_{k,i}}\right\} $
are assumed to be random variables which are mutually independent
and identically distributed (i.i.d.) and also independent of BS locations
$\left\{ \boldsymbol{\boldsymbol{X}_{k,i}}\right\} $, thus $h_{\boldsymbol{X}_{k,i}}$
can be denoted as $h_{k}$ for the sake of simplicity. Similarly,
the shadowing is denoted by $g_{k}$ and particularly assume that
it follows a log-normal distribution with zero mean and standard deviation
$\sigma$. Note that the proposed model is general enough to account
for various propagation scenarios with fasting fading, shadowing,
and different path loss models. 

To characterize shadowing effect in urban areas which is a unique
scenario in our analysis, both NLoS and LoS transmissions are incorporated.
That is, if the visual path between a BS \textbf{$\boldsymbol{X}_{k,i}\in\Phi_{k}$}
and the typical MU is blocked by obstacles like buildings, trees\lyxadded{Bean}{Thu Dec  7 09:55:12 2017}{,}
and even MUs, it is a\lyxadded{Bean}{Thu Dec  7 09:22:12 2017}{n}
NLoS transmission\lyxadded{Bean}{Thu Dec  7 09:54:15 2017}{.} \lyxadded{Bean}{Thu Dec  7 09:54:29 2017}{O}therwise
it is a LoS transmission. The occurrence of NLoS and LoS transmissions
depend on various environmental factors, including geographical structure,
distance\lyxadded{Bean}{Thu Dec  7 09:54:53 2017}{,} and cluster.
In this work, a one-parameter distance-based NLoS/LoS transmission
probability model is applied. That is,
\begin{equation}
p_{k}^{\textrm{NL}}\left(\left\Vert \boldsymbol{X}_{k,i}\right\Vert \right)+p_{k}^{\textrm{L}}\left(\left\Vert \boldsymbol{X}_{k,i}\right\Vert \right)=1,
\end{equation}
where $p_{k}^{\textrm{NL}}\left(\left\Vert \boldsymbol{X}_{k,i}\right\Vert \right)$
and $p_{k}^{\textrm{L}}\left(\left\Vert \boldsymbol{X}_{k,i}\right\Vert \right)$
denote the probability of the occurrence of NLoS and LoS transmissions,
respectively, $\left\Vert \boldsymbol{X}_{k,i}\right\Vert $ is the
distance between the BS \textbf{$\boldsymbol{X}_{k,i}$} and the typical
MU.

Regarding the mathematical form of $p_{k}^{\textrm{L}}\left(\left\Vert \boldsymbol{X}_{k,i}\right\Vert \right)$
(or $p_{k}^{\textrm{NL}}\left(\left\Vert \boldsymbol{X}_{k,i}\right\Vert \right)$),
Blaunstein \emph{et al.} \cite{Blaunstein98Parametric} formulated
$p_{k}^{\textrm{L}}\left(\left\Vert \boldsymbol{X}_{k,i}\right\Vert \right)$
as a negative exponential function, \emph{i.e.}, $p_{k}^{\textrm{L}}\left(\left\Vert \boldsymbol{X}_{k,i}\right\Vert \right)=e^{-\kappa\left\Vert \boldsymbol{X}_{k,i}\right\Vert }$,
where $\kappa$ is a parameter determined by the density and the mean
length of the blockages lying in the visual path between BSs and the
typical MU. Bai \emph{et al.} \cite{Bai14Analysis} extended Blaunstein's
work by using random shape theory which shows that $\kappa$ is not
only determined by the mean length but also the mean width of the
blockages. \cite{DiRenzo15StochasticJ} and \cite{Bai15Coverage}
approximated $p_{k}^{\textrm{L}}\left(\left\Vert \boldsymbol{X}_{k,i}\right\Vert \right)$
by using piece-wise functions and step functions, respectively. Ding
\emph{et al.} \cite{Ding16Performance} considered $p_{k}^{\textrm{L}}\left(\left\Vert \boldsymbol{X}_{k,i}\right\Vert \right)$
to be a linear function and a two-piece exponential function, respectively;
both are recommended by the 3GPP. It is important to note that the
introduction of NLoS and LoS transmissions is essential to model practical
networks, where a MU does not necessarily have to connect to the nearest
BS. Instead, for many cases, MUs are associated with farther BSs with
stronger signal strength. 

It should be noted that the occurrence of NLoS and LoS transmissions
is assumed to be independent for different BS-MU pairs. Though such
assumption might not be entirely realistic (\emph{e.g.}, NLoS transmission
caused by a large obstacle may be spatially correlated), Bai \emph{et
al.} \cite{Bai14Analysis,Bai15Coverage} showed that the impact of
the independence assumption on the SINR analysis is negligible. 

For a specific tier $k$, note that from the viewpoint of the typical
MU, each BS in the infinite plane $\mathbb{R}^{2}$ is either a\lyxadded{Bean}{Thu Dec  7 09:22:20 2017}{n}
NLoS BS or a LoS BS to the typical MU. Accordingly, a thinning procedure
on points in the PPP $\Phi_{k}$ is performed to model the distributions
of NLoS BSs and LoS BSs, respectively. That is, each BS in $\Phi_{k}$
will be kept if a BS has a\lyxadded{Bean}{Thu Dec  7 09:22:26 2017}{n}
NLoS transmission with the typical MU, thus forming a new point process
denoted by $\Phi_{k}^{\textrm{NL}}$ . While BSs in $\Phi_{k}\setminus\Phi_{k}^{\textrm{NL}}$
form another point process denoted by $\Phi_{k}^{\textrm{L}}$, representing
the set of BSs with LoS path to the typical MU. As a consequence of
the independence assumption of NLoS and LoS transmissions mentioned
in the last paragraph, $\Phi_{k}^{\textrm{NL}}$ and $\Phi_{k}^{\textrm{L}}$
are two independent non-homogeneous PPPs with intensity functions
$\lambda_{k}p_{k}^{\textrm{NL}}\left(\left\Vert \boldsymbol{X}_{k,i}\right\Vert \right)$
and $\lambda_{k}p_{k}^{\textrm{L}}\left(\left\Vert \boldsymbol{X}_{k,i}\right\Vert \right)$,
respectively.

Based on assumptions above, the received power of the typical MU from
a BS \textbf{$\boldsymbol{X}_{k,i}\in\Phi_{k}$} is defined as follows.

\lyxadded{Bean}{Mon Nov 20 06:36:54 2017}{\textbf{\hspace{-0.4cm}Definitation
1.} }The received power of the typical MU from a BS \textbf{$\boldsymbol{X}_{k,i}\in\Phi_{k}$}
, \emph{i.e.}, $P_{k,i}^{\textrm{rec}}$ is

\noindent 
\begin{equation}
P_{k,i}^{\textrm{rec}}=\begin{cases}
P_{k,i}^{\textrm{NL}}=P_{k}A_{k}^{\textrm{NL}}h_{k}^{\textrm{NL}}g_{k}^{\textrm{NL}}l_{k}^{\textrm{NL}}\left(\left\Vert \boldsymbol{X}_{k,i}\right\Vert \right), & \hspace{-0.3cm}\textrm{with probability }p_{k}^{\textrm{NL}}\left(\left\Vert \boldsymbol{X}_{k,i}\right\Vert \right)\\
P_{k,i}^{\textrm{L}}=P_{k}A_{k}^{\textrm{L}}h_{k}^{\textrm{L}}g_{k}^{\textrm{L}}l_{k}^{\textrm{L}}\left(\left\Vert \boldsymbol{X}_{k,i}\right\Vert \right), & \hspace{-0.3cm}\textrm{with probability }p_{k}^{\textrm{L}}\left(\left\Vert \boldsymbol{X}_{k,i}\right\Vert \right)
\end{cases},\label{eq:P_rec}
\end{equation}
\lyxadded{Bean}{Fri Nov 17 07:27:56 2017}{where $A_{k}^{\textrm{NL}}$
and $A_{k}^{\textrm{L}}$ denote the respective path loss for NLoS
and LoS transmissions at the reference distance (usually at 1 meter).}
For simplicity, denote $B_{k}^{\textrm{U}}=P_{k}A_{k}^{\textrm{U}}$
and let $l_{k}^{\textrm{U}}\left(\left\Vert \boldsymbol{X}_{k,i}\right\Vert \right)=\left\Vert \boldsymbol{X}_{k,i}\right\Vert ^{-\alpha_{k}^{\textrm{U}}}$,
where the superscript $\textrm{U}\in\left\{ \textrm{NL},\textrm{L}\right\} \overset{\textrm{def}}{=}\mathcal{U}$
used distinguishes NLoS and LoS transmissions and $\alpha_{k}^{\textrm{U}}$
denotes the path loss exponent for NLoS or LoS transmission in the
$k$-th tier. Recently, \cite{Liu17Effect} and \cite{AlAmmouri17SINR}
took bounded path loss model and stretched exponential path loss model
into consideration, in which several interesting performance trends
are found and will be investigated in our future work.

\lyxadded{Bean}{Mon Nov 20 06:38:35 2017}{\textbf{\hspace{-0.4cm}}\emph{Remark
}1. }Apart from the fixed transmit power, a density-dependent transmit
power is further assumed and analyzed mentioned in \cite{Ding17Performance},
\emph{i.e.}, $P_{k}\left(\lambda\right)=\frac{10^{\frac{T_{k}}{10}\eta}}{A_{k}^{\textrm{NL}}r_{k}^{-\alpha_{k}^{\textrm{NL}}}}$,
where $r_{k}=\sqrt{\frac{1}{\pi\lambda_{k}}}$ is the radius of an
equivalent disk-shaped coverage area in the $k$-th tier with an area
size of $r_{k}=\sqrt{\frac{1}{\pi}}$\lyxadded{Bean}{Fri Nov 17 07:37:53 2017}{
and $T_{k}$ is the per tier SINR threshold}. 

\subsection{Cell Association Scheme}

Cell association scheme\lyxadded{Bean}{Thu Dec  7 10:52:13 2017}{
\cite{Yang15A}} plays a \lyxadded{Bean}{Thu Dec  7 09:57:07 2017}{crucial}
role in network performance determining BS coverage, MU hand-off regulation
and even facility deployment of small cells. Conventionally, a typical
MU is connected to the BS $\boldsymbol{X}_{k,m}$ if and only if 

\begin{equation}
\mathcal{\mathcal{P}}_{k,m}^{\textrm{dBm}}>\mathcal{\mathcal{P}}_{j,n}^{\textrm{dBm}},j\neq k,\label{eq:MIRPEQ}
\end{equation}
where $\mathcal{\mathcal{P}}_{k,m}^{\textrm{dBm}}$ is the instantaneous
received power with dBm unit from the BS $\boldsymbol{X}_{k,m}$ and
Eq. (\ref{eq:MIRPEQ}) is known as the MIRP association scheme.

In practical, $\mathcal{\mathcal{P}}_{k}^{\textrm{dBm}}$ is usually
averaged out in time and frequency domains to cope with fluctuations
caused by channel fading. In this text, a typical MU is connected
to the BS $\boldsymbol{X}_{k,m}$ if and only if 
\begin{equation}
\overline{\mathcal{\mathcal{P}}_{k,m}^{\textrm{dBm}}}>\overline{\mathcal{\mathcal{P}}_{j,n}^{\textrm{dBm}}},j\neq k,\label{eq:MARPEQ}
\end{equation}
where $\overline{\mathcal{\mathcal{P}}_{k,m}^{\textrm{dBm}}}$ denotes
the average received power with dBm unit from the BS $\boldsymbol{X}_{k,m}$
and Eq. (\ref{eq:MARPEQ}) is known as the MARP association scheme.

Aided by \emph{cell range expansion} (CRE), which is realized by MUs
adding a positive \emph{cell range expansion bias} (CREB) to the received
power from BSs in different tiers, more MUs can be offloaded to small
cells. That is, if a MU is associated with the BS $\boldsymbol{X}_{k,m}$
if and only if

\begin{equation}
\overline{\mathcal{\mathcal{P}}_{k,m}^{\textrm{dBm}}}+\triangle_{k,m}^{\textrm{dB}}>\overline{\mathcal{\mathcal{P}}_{j,n}^{\textrm{dBm}}}+\triangle_{j.n}^{\textrm{dB}},j\neq k,
\end{equation}
where $\triangle_{k,m}^{\textrm{dB}}$ and $\triangle_{j,n}^{\textrm{dB}}$is
the CREB with dB unit in the $k$-th and $j$-th tier. With proper
CREB chosen, the coverage of BSs in some tiers is artificially expanded,
allowing MUs more \lyxadded{Bean}{Thu Dec  7 09:58:47 2017}{flexible}
to be associated with BSs which may not provide the strongest received
power, thus balancing traffic load to achieve spatial efficiency.
However, CRE causes severe interference to small cell MU which impair
the QoS of small cell users and thus \emph{almost blank subframes}
(ABS) coordination is needed between macrocell BSs and small cell
BSs. However, the analysis of CRE plus ABS is challenging because
(i) the association scheme is not only determined by the received
power but also the current resource allocation strategy, and (ii)
ignoring ABS while using CRE can impair the coverage performance.
For simplicity, CRE and ABS are not going to be considered in this
paper, which are left as our future work. 

\subsection{Performance Metrics}

To evaluate the network performance, the following three metrics,
\emph{i.e.}, the coverage probability, the PT and the EE, are focused
on.

The coverage probability is the probability that the received SINR
is greater than a given threshold, i,e, $p_{\textrm{cov}}\left(\left\{ \lambda_{k}\right\} ,\left\{ T_{k}\right\} ,\left\{ B_{k}^{\textrm{U}}\right\} \right)=\Pr\Bigl[\underset{k\in\mathcal{K},\boldsymbol{X}_{k,i}\in\Phi_{k}}{\cup}\textrm{SINR}_{k}\left(\left\Vert \boldsymbol{X}_{k,i}\right\Vert \right)>T_{k}\Bigr]$,
where $\textrm{SINR}_{k}\left(\left\Vert \boldsymbol{X}_{k,i}\right\Vert \right)$
is defined as follows
\begin{equation}
\textrm{SINR}_{k}\left(\left\Vert \boldsymbol{X}_{k,i}\right\Vert \right)=\frac{P_{k}A_{k}^{\textrm{U}}h_{k}^{\textrm{U}}g_{k}^{\textrm{U}}l_{k}^{\textrm{U}}\left(\left\Vert \boldsymbol{X}_{k,i}\right\Vert \right)}{\stackrel[k=1]{K}{\sum}\underset{\boldsymbol{X}_{k,j}\in\Phi_{k}\setminus\boldsymbol{X}_{k,i}}{\sum}P_{k}A_{k}^{\textrm{U}}h_{k}^{\textrm{U}}g_{k}^{\textrm{U}}l_{k}^{\textrm{U}}\left(\left\Vert \boldsymbol{X}_{k,j}\right\Vert \right)+\eta},
\end{equation}
where $\Phi_{k}\setminus\boldsymbol{X}_{k,i}$ is the Palm point process~\cite{Chiu13Stochastic}
representing the set of interfering BSs in the $k$-th tier and $\eta$
denotes the noise power at the MU side, which is assumed to be the
additive white Gaussian noise (AWGN).

The PT is defined as follows \cite{Bao14Structured,AlAmmouri17SINR}
\begin{equation}
\mathcal{T}\left(\left\{ \lambda_{k}\right\} ,\left\{ T_{k}\right\} ,\left\{ B_{k}^{\textrm{U}}\right\} \right)=\stackrel[k=1]{K}{\sum}\lambda_{k}\mathcal{A}_{k}p_{\textrm{cov},k}^{\textrm{cond}}\log_{2}\left(1+T_{k}\right)=\stackrel[k=1]{K}{\sum}\lambda_{k}p_{\textrm{cov},k}\log_{2}\left(1+T_{k}\right),\label{eq:Throughput}
\end{equation}
where the network is fully loaded due to the assumption that $\lambda_{u}\gg\lambda_{k}$,
$\mathcal{A}_{k}$ is the association probability that the typical
MU is connected to the $k$-th tier, $p_{\textrm{cov},k}^{\textrm{cond}}$
is the conditional association coverage probability and $p_{\textrm{cov},k}$
is the per-tier coverage probability. Compared with the area spectral
efficiency (ASE), which is defined as 
\begin{equation}
\textrm{ASE}\left(\left\{ \lambda_{k}\right\} ,\left\{ T_{k}\right\} ,\left\{ B_{k}^{\textrm{U}}\right\} \right)=\stackrel[k=1]{K}{\sum}\mathbb{E}\left[\lambda_{k}\log_{2}\left(1+\textrm{SINR}_{k}\left(\left\Vert \boldsymbol{X}_{k,i}\right\Vert \right)\right)\right],
\end{equation}
the PT implicitly assumes a fixed rate transmission from all BSs in
the network, and has a unit of $\textrm{bps/Hz/m}^{2}$, while the
ASE assumes full buffers but it allows each link to adapt its rate
to the optimal value for a given SINR, thus avoiding outages at low
SINR and the wasting of rate at high SINR \cite{AlAmmouri17SINR}.
In other words, the PT is a more realistic performance metric and
the ASE upper bounds the PT. In our analysis, the PT is chosen as
our performance metric.

The EE is defined as the ratio between the PT and the total energy
consumption of the network, \emph{i.e.},
\begin{equation}
\mathcal{E}\left(\left\{ \lambda_{k}\right\} ,\left\{ T_{k}\right\} ,\left\{ B_{k}^{\textrm{U}}\right\} \right)=\frac{\mathcal{T}\left(\left\{ \lambda_{k}\right\} ,\left\{ T_{k}\right\} ,\left\{ B_{k}^{\textrm{U}}\right\} \right)}{\stackrel[k=1]{K}{\sum}\lambda_{k}\left(a_{k}P_{k}+b_{k}\right)},\label{eq:EE}
\end{equation}
where the coefficient $a_{k}$ accounts for power consumption that
scales with the average radiated power, and the term $b_{k}$ models
the static power consumed by signal processing, battery backup and
cooling \cite{Peng15Energy}. Other performance metrics, such as the
bit-error probability and per-MU data rate, can be found using the
coverage probability (SINR distribution) following the methods mentioned
in \cite{ElSawy17Modeling}.

\section{\label{sec:Performance Analysis} Performance Analysis}

In this section, we derive expressions for the considered performance
metrics and study the effect of densification on these metrics. It
is started by introducing the network transformation and then presenting
the analytical expressions with the MIRP and MARP association schemes
in the following subsections. 

\subsection{Network Transformation}

Before presenting our main analytical results, firstly the network
transformation is introduced, which aims to unify the analysis and
to reduce the complexity as well. 

Using the manipulation in \cite{Yang17Density,Ge15Spatial}, we define
\begin{equation}
\overline{R_{k,i}^{\textrm{NL}}}=\left\Vert \boldsymbol{X}_{k,i}\right\Vert \cdot\left(B_{k}^{\textrm{NL}}g_{k}^{\textrm{NL}}\right)^{-1/\alpha_{k}^{\textrm{NL}}},\label{eq:R_N}
\end{equation}
and
\begin{equation}
\overline{R_{k,i}^{\textrm{L}}}=\left\Vert \boldsymbol{X}_{k,i}\right\Vert \cdot\left(B_{k}^{\textrm{L}}g_{k}^{\textrm{L}}\right)^{-1/\alpha_{k}^{\textrm{L}}},\label{eq:R_L}
\end{equation}
respectively. Then Eq. (\ref{eq:P_rec}) can be written as
\begin{equation}
P_{k,i}^{\textrm{rec}}=\begin{cases}
P_{k,i}^{\textrm{NL}}=h_{k}^{\textrm{NL}}\left(\overline{R_{k,i}^{\textrm{NL}}}\right)^{-\alpha_{k}^{\textrm{NL}}}, & \hspace{-0.3cm}\textrm{with probability }p_{k}^{\textrm{NL}}\left(\left\Vert \boldsymbol{X}_{k,i}\right\Vert \right)\\
P_{k,i}^{\textrm{L}}=h_{k}^{\textrm{L}}\left(\overline{R_{k,i}^{\textrm{L}}}\right)^{-\alpha_{k}^{\textrm{L}}}, & \hspace{-0.3cm}\textrm{with probability }p_{k}^{\textrm{L}}\left(\left\Vert \boldsymbol{X}_{k,i}\right\Vert \right)
\end{cases}.\label{eq:P_rec_equivalence}
\end{equation}

By adopting the Equivalence Theorem in \cite{Yang17Density}, it is
concluded that the distance $\left\{ \overline{R_{k,i}^{\textrm{NL}}}\right\} _{i}$
(or $\left\{ \overline{R_{k,i}^{\textrm{L}}}\right\} _{i}$) \lyxadded{Bean}{Fri Nov 17 08:47:12 2017}{from}
a scaled point process for NLoS BSs (or LoS BSs), which still remains
a PPP denoted by $\overline{\Phi_{k}^{\textrm{NL}}}$ (or $\overline{\Phi_{k}^{\textrm{L}}}$
). $\left\{ \overline{\Phi_{k}^{\textrm{U}}}\right\} _{k},\textrm{U}\in\mathcal{U}$
are mutually independent with each other\lyxadded{Bean}{Thu Dec  7 10:02:55 2017}{,}
and the intensity measures and intensities are provided in Lemma 1
as below.

\lyxadded{Bean}{Mon Nov 20 06:40:10 2017}{\textbf{\hspace{-0.4cm}Lemma
1.}} The intensity measure and intensity of $\overline{\Phi_{k}^{\textrm{U}}}$
can be formulated as
\begin{equation}
\lambda_{k}^{\textrm{NL}}\left(t\right)=\frac{\textrm{d}}{\textrm{d}t}\Lambda_{k}^{\textrm{NL}}\left(\left[0,t\right]\right),\label{eq:Lambda_N_k}
\end{equation}
and
\begin{equation}
\lambda_{k}^{\textrm{L}}\left(t\right)=\frac{\textrm{d}}{\textrm{d}t}\Lambda_{k}^{\textrm{L}}\left(\left[0,t\right]\right),\label{eq:Lambda_L_k}
\end{equation}
respectively, where
\begin{equation}
\Lambda_{k}^{\textrm{NL}}\left(\left[0,t\right]\right)=\mathbb{E}_{g_{k}^{\textrm{NL}}}\Bigl[2\pi\lambda_{k}\int_{z=0}^{t\left(B_{k}^{\textrm{NL}}g_{k}^{\textrm{NL}}\right)^{1/\alpha_{k}^{\textrm{NL}}}}p_{k}^{\textrm{NL}}\left(z\right)z\textrm{d}z\Bigr]\label{eq:Measure_N_k}
\end{equation}
and
\begin{equation}
\Lambda_{k}^{\textrm{L}}\left(\left[0,t\right]\right)=\mathbb{E}_{g_{k}^{\textrm{L}}}\Bigl[2\pi\lambda_{k}\int_{z=0}^{t\left(B_{k}^{\textrm{L}}g_{k}^{\textrm{L}}\right)^{1/\alpha_{k}^{\textrm{L}}}}p_{k}^{\textrm{L}}\left(z\right)z\textrm{d}z\Bigr].\label{eq:Measure_L_k}
\end{equation}
\begin{IEEEproof}
The proof can be referred to \cite[Appendix A]{Yang17Density} and
thus omitted here. Aided by the network transformation and stochastic
geometry tool, the coverage probability, the PT and the EE will be
derived in the following.
\end{IEEEproof}

\subsection{\label{subsec:Pc MIRP}Coverage Probability with the MIRP Association
Scheme}

With the MIRP association scheme, the typical MU is associated with
the BS which offers the maximum instantaneous received power as shown
in Eq. (\ref{eq:MIRPEQ}). Using this cell association scheme and
considering Lemma 1, the general results of coverage probability in
the $K$-tier HetNets is given by Theorem 1.

\lyxadded{Bean}{Mon Nov 20 06:43:32 2017}{\textbf{\hspace{-0.4cm}Theorem
1.} }When $T_{k}\geqslant1$, the coverage probability for a typical
MU with the MIRP association scheme can be derived as 
\begin{align}
 & \quad\,p_{\textrm{cov}}^{\textrm{MIRP}}\left(\left\{ \lambda_{k}\right\} ,\left\{ T_{k}\right\} ,\left\{ B_{k}^{\textrm{U}}\right\} \right)\nonumber \\
 & =\stackrel[k=1]{K}{\sum}\int_{r=0}^{\infty}e^{-T_{k}\eta r^{\alpha_{k}^{\textrm{NL}}}}\lambda_{k}^{\textrm{NL}}\left(r\right)\stackrel[j=1]{K}{\prod}\left[\mathcal{L}_{I_{j}^{\textrm{NL}}}^{\textrm{MIRP}}\left(T_{k}r^{\alpha_{k}^{\textrm{NL}}}\right)\mathcal{L}_{I_{j}^{\textrm{L}}}^{\textrm{MIRP}}\left(T_{k}r^{\alpha_{k}^{\textrm{NL}}}\right)\right]\textrm{d}r\nonumber \\
 & \quad\,+\stackrel[k=1]{K}{\sum}\int_{t=0}^{\infty}e^{-T_{k}\eta r^{\alpha_{k}^{\textrm{L}}}}\lambda_{k}^{\textrm{L}}\left(r\right)\stackrel[j=1]{K}{\prod}\left[\mathcal{L}_{I_{j}^{\textrm{NL}}}^{\textrm{MIRP}}\left(T_{k}r^{\alpha_{k}^{\textrm{L}}}\right)\mathcal{L}_{I_{j}^{\textrm{L}}}^{\textrm{MIRP}}\left(T_{k}r^{\alpha_{k}^{\textrm{L}}}\right)\right]\textrm{d}r,\label{eq:PC_MIRP}
\end{align}
where 
\begin{equation}
\mathcal{L}_{I_{j}^{\textrm{NL}}}^{\textrm{MIRP}}\left(s\right)=\exp\Bigl[-\int_{y=0}^{\infty}\frac{\lambda_{j}^{\textrm{NL}}\left(y\right)}{1+y^{\alpha_{j}^{\textrm{NL}}}/s}\textrm{d}y\Bigr],\label{eq:LT_NL_max_ins}
\end{equation}
and
\begin{equation}
\mathcal{L}_{I_{j}^{\textrm{L}}}^{\textrm{MIRP}}\left(s\right)=\exp\Bigl[-\int_{y=0}^{\infty}\frac{\lambda_{j}^{\textrm{L}}\left(y\right)}{1+y^{\alpha_{j}^{\textrm{L}}}/s}\textrm{d}y\Bigr].\label{eq:LT_L_max_ins}
\end{equation}
\begin{IEEEproof}
See Appendix A.
\end{IEEEproof}
In pursuit of the analytical results of the PT and the EE, the NLoS/LoS
coverage probability and per-tier coverage probability are presented
in the following two corollaries.

\lyxadded{Bean}{Mon Nov 20 07:23:26 2017}{\textbf{\hspace{-0.4cm}Corollary
1.} }When $T_{k}\geqslant1$, the coverage probability for a typical
MU which is served by NLoS BSs and LoS BSs with the MIRP association
scheme are given by
\begin{align}
 & \quad\,p_{\textrm{cov,NL}}^{\textrm{MIRP}}\left(\left\{ \lambda_{k}\right\} ,\left\{ T_{k}\right\} ,\left\{ B_{k}^{\textrm{U}}\right\} \right)=\stackrel[k=1]{K}{\sum}p_{\textrm{NL},k}^{\textrm{MIRP}}\left(\left\{ \lambda_{k}\right\} ,\left\{ T_{k}\right\} ,\left\{ B_{k}^{\textrm{U}}\right\} \right)\nonumber \\
 & =\stackrel[k=1]{K}{\sum}\int_{r=0}^{\infty}e^{-T_{k}\eta r^{\alpha_{k}^{\textrm{NL}}}}\lambda_{k}^{\textrm{NL}}\left(r\right)\stackrel[j=1]{K}{\prod}\left[\mathcal{L}_{I_{j}^{\textrm{NL}}}^{\textrm{MIRP}}\left(T_{k}r^{\alpha_{k}^{\textrm{NL}}}\right)\mathcal{L}_{I_{j}^{\textrm{L}}}^{\textrm{MIRP}}\left(T_{k}r^{\alpha_{k}^{\textrm{NL}}}\right)\right]\textrm{d}r,
\end{align}
and 
\begin{align}
 & \quad\,p_{\textrm{cov,L}}^{\textrm{MIRP}}\left(\left\{ \lambda_{k}\right\} ,\left\{ T_{k}\right\} ,\left\{ B_{k}^{\textrm{U}}\right\} \right)=\stackrel[k=1]{K}{\sum}p_{\textrm{L},k}^{\textrm{MIRP}}\left(\left\{ \lambda_{k}\right\} ,\left\{ T_{k}\right\} ,\left\{ B_{k}^{\textrm{U}}\right\} \right)\nonumber \\
 & =\stackrel[k=1]{K}{\sum}\int_{t=0}^{\infty}e^{-T_{k}\eta r^{\alpha_{k}^{\textrm{L}}}}\lambda_{k}^{\textrm{L}}\left(r\right)\stackrel[j=1]{K}{\prod}\left[\mathcal{L}_{I_{j}^{\textrm{NL}}}^{\textrm{MIRP}}\left(T_{k}r^{\alpha_{k}^{\textrm{L}}}\right)\mathcal{L}_{I_{j}^{\textrm{L}}}^{\textrm{MIRP}}\left(T_{k}r^{\alpha_{k}^{\textrm{L}}}\right)\right]\textrm{d}r,
\end{align}
respectively.
\begin{IEEEproof}
This corollary can be derived \lyxadded{Bean}{Fri Nov 17 03:18:07 2017}{from}
Theorem 1 by rearranging the terms in Eq. (\ref{eq:PC_MIRP}) and
thus the proof is omitted here.
\end{IEEEproof}
\lyxadded{Bean}{Mon Nov 20 07:24:16 2017}{\textbf{\hspace{-0.4cm}Corollary
2.} }When $T_{k}\geqslant1$, the per-tier coverage probability for
a typical MU which is covered by the $k$-th tier with the MIRP association
scheme is given by
\begin{align}
 & \quad\,p_{\textrm{cov},k}^{\textrm{MIRP}}\left(\left\{ \lambda_{k}\right\} ,\left\{ T_{k}\right\} ,\left\{ B_{k}^{\textrm{U}}\right\} \right)\nonumber \\
 & =\int_{r=0}^{\infty}e^{-T_{k}\eta r^{\alpha_{k}^{\textrm{NL}}}}\lambda_{k}^{\textrm{NL}}\left(r\right)\stackrel[j=1]{K}{\prod}\left[\mathcal{L}_{I_{j}^{\textrm{NL}}}^{\textrm{MIRP}}\left(T_{k}r^{\alpha_{k}^{\textrm{NL}}}\right)\mathcal{L}_{I_{j}^{\textrm{L}}}^{\textrm{MIRP}}\left(T_{k}r^{\alpha_{k}^{\textrm{NL}}}\right)\right]\textrm{d}r\nonumber \\
 & \quad\,+\int_{t=0}^{\infty}e^{-T_{k}\eta r^{\alpha_{k}^{\textrm{L}}}}\lambda_{k}^{\textrm{L}}\left(r\right)\stackrel[j=1]{K}{\prod}\left[\mathcal{L}_{I_{j}^{\textrm{NL}}}^{\textrm{MIRP}}\left(T_{k}r^{\alpha_{k}^{\textrm{L}}}\right)\mathcal{L}_{I_{j}^{\textrm{L}}}^{\textrm{MIRP}}\left(T_{k}r^{\alpha_{k}^{\textrm{L}}}\right)\right]\textrm{d}r.
\end{align}
\begin{IEEEproof}
This corollary can be derived \lyxadded{Bean}{Fri Nov 17 08:47:38 2017}{from}
Theorem 1 by rearranging the terms in Eq. (\ref{eq:PC_MIRP}) and
thus the proof is omitted here.
\end{IEEEproof}

\subsection{Coverage Probability with the MARP Association Scheme}

With the MARP association scheme, the typical MU is associated with
the BS which offers the maximum long-term averaged received power
by averaging out the effect of multi-path fading $h_{k}^{\textrm{U}}$.
With this cell association scheme, the \lyxadded{Bean}{Thu Dec  7 10:04:45 2017}{primary}
results of coverage probability is given by Theorem 2.

\lyxadded{Bean}{Mon Nov 20 07:25:11 2017}{\textbf{\hspace{-0.4cm}Theorem
2.} }The coverage probability for a typical MU with the MARP association
scheme is
\begin{align}
 & \quad\,p_{\textrm{cov}}^{\textrm{MARP}}\left(\left\{ \lambda_{k}\right\} ,\left\{ T_{k}\right\} ,\left\{ B_{k}^{\textrm{U}}\right\} \right)\nonumber \\
 & =\stackrel[k=1]{K}{\sum}p_{k,\textrm{NL}}^{\textrm{MARP}}\left(\left\{ \lambda_{k}\right\} ,\left\{ T_{k}\right\} ,\left\{ B_{k}^{\textrm{U}}\right\} \right)+\stackrel[k=1]{K}{\sum}p_{k,\textrm{L}}^{\textrm{MARP}}\left(\left\{ \lambda_{k}\right\} ,\left\{ T_{k}\right\} ,\left\{ B_{k}^{\textrm{U}}\right\} \right)\nonumber \\
 & =\stackrel[k=1]{K}{\sum}\int_{r=0}^{\infty}e^{-T_{k}\eta r^{\alpha_{k}^{\textrm{NL}}}}\lambda_{k}^{\textrm{NL}}\left(r\right)\stackrel[j=1]{K}{\prod}\left[\mathcal{L}_{I_{j}^{\textrm{NL}}}^{\textrm{MARP1}}\left(T_{k}r^{\alpha_{k}^{\textrm{NL}}}\right)\mathcal{L}_{I_{j}^{\textrm{L}}}^{\textrm{MARP1}}\left(T_{k}r^{\alpha_{k}^{\textrm{NL}}}\right)\right]\nonumber \\
 & \quad\,\times e^{-\stackrel[j=1]{K}{\sum}\left[\Lambda_{j}^{\textrm{L}}\left(\left[0,r^{\alpha_{k}^{\textrm{NL}}/\alpha_{j}^{\textrm{L}}}\right]\right)+\Lambda_{j}^{\textrm{NL}}\left(\left[0,r^{\alpha_{k}^{\textrm{NL}}/\alpha_{j}^{\textrm{NL}}}\right]\right)\right]}\textrm{d}r\nonumber \\
 & \quad\,+\stackrel[k=1]{K}{\sum}\int_{r=0}^{\infty}e^{-T_{k}\eta r^{\alpha_{k}^{\textrm{L}}}}\lambda_{k}^{\textrm{L}}\left(r\right)\stackrel[j=1]{K}{\prod}\left[\mathcal{L}_{I_{j}^{\textrm{NL}}}^{\textrm{MARP2}}\left(T_{k}r^{\alpha_{k}^{\textrm{L}}}\right)\mathcal{L}_{I_{j}^{\textrm{L}}}^{\textrm{MARP2}}\left(T_{k}r^{\alpha_{k}^{\textrm{L}}}\right)\right]\nonumber \\
 & \quad\,\times e^{-\stackrel[j=1]{K}{\sum}\left[\Lambda_{j}^{\textrm{L}}\left(\left[0,r^{\alpha_{k}^{\textrm{L}}/\alpha_{j}^{\textrm{L}}}\right]\right)+\Lambda_{j}^{\textrm{NL}}\left(\left[0,r^{\alpha_{k}^{\textrm{L}}/\alpha_{j}^{\textrm{NL}}}\right]\right)\right]}\textrm{d}r,\label{eq:PC_MARP}
\end{align}
where 
\begin{equation}
\mathcal{L}_{I_{j}^{\textrm{NL}}}^{\textrm{MARP1}}\left(s\right)=\exp\Bigl[-\int_{y=r^{\alpha_{k}^{\textrm{NL}}/\alpha_{j}^{\textrm{NL}}}}^{\infty}\frac{\lambda_{j}^{\textrm{NL}}\left(y\right)}{1+y^{\alpha_{j}^{\textrm{NL}}}/s}\textrm{d}y\Bigr],\label{eq:LT_NL_max_ave1}
\end{equation}
\begin{equation}
\mathcal{L}_{I_{j}^{\textrm{L}}}^{\textrm{MARP1}}\left(s\right)=\exp\Bigl[-\int_{y=r^{\alpha_{k}^{\textrm{NL}}/\alpha_{j}^{\textrm{L}}}}^{\infty}\frac{\lambda_{j}^{\textrm{L}}\left(y\right)}{1+y^{\alpha_{j}^{\textrm{L}}}/s}\textrm{d}y\Bigr],\label{eq:LT_L_max_ave1}
\end{equation}
\begin{equation}
\mathcal{L}_{I_{j}^{\textrm{NL}}}^{\textrm{MARP2}}\left(s\right)=\exp\Bigl[-\int_{y=r^{\alpha_{k}^{\textrm{L}}/\alpha_{j}^{\textrm{NL}}}}^{\infty}\frac{\lambda_{j}^{\textrm{NL}}\left(y\right)}{1+y^{\alpha_{j}^{\textrm{NL}}}/s}\textrm{d}y\Bigr],\label{eq:LT_NL_max_ave2}
\end{equation}
and
\begin{equation}
\mathcal{L}_{I_{j}^{\textrm{L}}}^{\textrm{MARP2}}\left(s\right)=\exp\Bigl[-\int_{y=r^{\alpha_{k}^{\textrm{L}}/\alpha_{j}^{\textrm{L}}}}^{\infty}\frac{\lambda_{j}^{\textrm{L}}\left(y\right)}{1+y^{\alpha_{j}^{\textrm{L}}}/s}\textrm{d}y\Bigr].\label{eq:LT_L_max_ave2}
\end{equation}
\begin{IEEEproof}
See Appendix B.
\end{IEEEproof}
\lyxadded{Bean}{Thu Dec  7 10:05:47 2017}{\textbf{\hspace{-0.4cm}}\emph{Remark
}2. }Note that different \lyxadded{Bean}{Fri Nov 17 08:47:50 2017}{from}
Theorem 1, Theorem 2 can be applied to scenarios without the assumption
of a particular range of SINR threshold $T_{k}$, \emph{e.g.}, $T_{k}\geqslant1$.

Similar to the study for Theorem 1, we provide two corollaries, \emph{i.e.},
the NLoS/LoS coverage probability and the per-tier coverage probability,
as follows.

\lyxadded{Bean}{Mon Nov 20 07:27:06 2017}{\textbf{\hspace{-0.4cm}Corollary
3.} }The coverage probability for a typical MU which is served by
NLoS BSs and LoS BSs with the MARP association scheme are given by
\begin{align}
 & \quad\,p_{\textrm{cov,NL}}^{\textrm{MARP}}\left(\left\{ \lambda_{k}\right\} ,\left\{ T_{k}\right\} ,\left\{ B_{k}^{\textrm{U}}\right\} \right)=\stackrel[k=1]{K}{\sum}p_{\textrm{NL},k}^{\textrm{MARP}}\left(\left\{ \lambda_{k}\right\} ,\left\{ T_{k}\right\} ,\left\{ B_{k}^{\textrm{U}}\right\} \right)\nonumber \\
 & =\stackrel[k=1]{K}{\sum}\int_{r=0}^{\infty}e^{-T_{k}\eta r^{\alpha_{k}^{\textrm{NL}}}}\lambda_{k}^{\textrm{NL}}\left(r\right)\stackrel[j=1]{K}{\prod}\left[\mathcal{L}_{I_{j}^{\textrm{NL}}}^{\textrm{MARP1}}\left(T_{k}r^{\alpha_{k}^{\textrm{NL}}}\right)\mathcal{L}_{I_{j}^{\textrm{L}}}^{\textrm{MARP1}}\left(T_{k}r^{\alpha_{k}^{\textrm{NL}}}\right)\right]\nonumber \\
 & \quad\,\times e^{-\stackrel[j=1]{K}{\sum}\left[\Lambda_{j}^{\textrm{L}}\left(\left[0,r^{\alpha_{k}^{\textrm{NL}}/\alpha_{j}^{\textrm{L}}}\right]\right)+\Lambda_{j}^{\textrm{NL}}\left(\left[0,r^{\alpha_{k}^{\textrm{NL}}/\alpha_{j}^{\textrm{NL}}}\right]\right)\right]}\textrm{d}r
\end{align}
and 
\begin{align}
 & \quad\,p_{\textrm{cov,L}}^{\textrm{MARP}}\left(\left\{ \lambda_{k}\right\} ,\left\{ T_{k}\right\} ,\left\{ B_{k}^{\textrm{U}}\right\} \right)=\stackrel[k=1]{K}{\sum}p_{\textrm{L},k}^{\textrm{MARP}}\left(\left\{ \lambda_{k}\right\} ,\left\{ T_{k}\right\} ,\left\{ B_{k}^{\textrm{U}}\right\} \right)\nonumber \\
 & =\stackrel[k=1]{K}{\sum}\int_{r=0}^{\infty}e^{-T_{k}\eta r^{\alpha_{k}^{\textrm{L}}}}\lambda_{k}^{\textrm{L}}\left(r\right)\stackrel[j=1]{K}{\prod}\left[\mathcal{L}_{I_{j}^{\textrm{NL}}}^{\textrm{MARP2}}\left(T_{k}r^{\alpha_{k}^{\textrm{L}}}\right)\mathcal{L}_{I_{j}^{\textrm{L}}}^{\textrm{MARP2}}\left(T_{k}r^{\alpha_{k}^{\textrm{L}}}\right)\right]\nonumber \\
 & \quad\,\times e^{-\stackrel[j=1]{K}{\sum}\left[\Lambda_{j}^{\textrm{L}}\left(\left[0,r^{\alpha_{k}^{\textrm{L}}/\alpha_{j}^{\textrm{L}}}\right]\right)+\Lambda_{j}^{\textrm{NL}}\left(\left[0,r^{\alpha_{k}^{\textrm{L}}/\alpha_{j}^{\textrm{NL}}}\right]\right)\right]}\textrm{d}r,
\end{align}
respectively.
\begin{IEEEproof}
This corollary can be derived \lyxadded{Bean}{Fri Nov 17 08:48:02 2017}{from}
Theorem 2 by rearranging the terms in Eq. (\ref{eq:PC_MARP}) and
thus the proof is omitted here.
\end{IEEEproof}
\lyxadded{Bean}{Mon Nov 20 07:27:18 2017}{\textbf{\hspace{-0.4cm}Corollary
4.} }The per-tier coverage probability for a typical MU which is covered
by the $k$-th tier with the MARP association scheme is given by
\begin{align*}
 & \quad\,p_{\textrm{cov},k}^{\textrm{MARP}}\left(\left\{ \lambda_{k}\right\} ,\left\{ T_{k}\right\} ,\left\{ B_{k}^{\textrm{U}}\right\} \right)
\end{align*}
\begin{align}
 & =\int_{r=0}^{\infty}e^{-T_{k}\eta r^{\alpha_{k}^{\textrm{NL}}}}\lambda_{k}^{\textrm{NL}}\left(r\right)\stackrel[j=1]{K}{\prod}\left[\mathcal{L}_{I_{j}^{\textrm{NL}}}^{\textrm{MARP1}}\left(T_{k}r^{\alpha_{k}^{\textrm{NL}}}\right)\mathcal{L}_{I_{j}^{\textrm{L}}}^{\textrm{MARP1}}\left(T_{k}r^{\alpha_{k}^{\textrm{NL}}}\right)\right]\nonumber \\
 & \quad\,\times e^{-\stackrel[j=1]{K}{\sum}\left[\Lambda_{j}^{\textrm{L}}\left(\left[0,r^{\alpha_{k}^{\textrm{NL}}/\alpha_{j}^{\textrm{L}}}\right]\right)+\Lambda_{j}^{\textrm{NL}}\left(\left[0,r^{\alpha_{k}^{\textrm{NL}}/\alpha_{j}^{\textrm{NL}}}\right]\right)\right]}\textrm{d}r\nonumber \\
 & \quad\,+\int_{r=0}^{\infty}e^{-T_{k}\eta r^{\alpha_{k}^{\textrm{L}}}}\lambda_{k}^{\textrm{L}}\left(r\right)\stackrel[j=1]{K}{\prod}\left[\mathcal{L}_{I_{j}^{\textrm{NL}}}^{\textrm{MARP2}}\left(T_{k}r^{\alpha_{k}^{\textrm{L}}}\right)\mathcal{L}_{I_{j}^{\textrm{L}}}^{\textrm{MARP2}}\left(T_{k}r^{\alpha_{k}^{\textrm{L}}}\right)\right]\nonumber \\
 & \quad\,\times e^{-\stackrel[j=1]{K}{\sum}\left[\Lambda_{j}^{\textrm{L}}\left(\left[0,r^{\alpha_{k}^{\textrm{L}}/\alpha_{j}^{\textrm{L}}}\right]\right)+\Lambda_{j}^{\textrm{NL}}\left(\left[0,r^{\alpha_{k}^{\textrm{L}}/\alpha_{j}^{\textrm{NL}}}\right]\right)\right]}\textrm{d}r.
\end{align}
\begin{IEEEproof}
This corollary can be derived \lyxadded{Bean}{Fri Nov 17 08:48:28 2017}{from}
Theorem 2 by rearranging the terms in Eq. (\ref{eq:PC_MARP}) and
thus the proof is omitted here.
\end{IEEEproof}
\lyxadded{Bean}{Thu Dec  7 10:06:27 2017}{Intuitively, the coverage
probability with the MIRP association scheme is higher than that with
the MARP association scheme. However, it can be proved mathematically
which is summarized in the following corollary. }

\lyxadded{Bean}{Thu Dec  7 10:07:50 2017}{\textbf{\hspace{-0.4cm}Corollary
5.} In the studied $K$-tier HetNet, the coverage probability with
the MIRP association scheme is higher than that with the MARP association
scheme, where the gap is determined by the intensity and the intensity
measure.}
\begin{IEEEproof}
\lyxadded{Bean}{Wed Dec  6 12:39:14 2017}{See Appendix C.}
\end{IEEEproof}

\subsection{The PT and the EE}

As the results with the MIRP and the MARP association schemes are
some kind of similar and the MARP association scheme is more practical
in the real network, we take the MARP association scheme as an example
to evaluate the PT and the EE in the following. The PT with the MARP
association scheme can be directly obtained from the coverage probability
expressions using Eq. (\ref{eq:Throughput}), \emph{i.e.}, 
\begin{equation}
\mathcal{T}\left(\left\{ \lambda_{k}\right\} ,\left\{ T_{k}\right\} ,\left\{ B_{k}^{\textrm{U}}\right\} \right)=\stackrel[k=1]{K}{\sum}\lambda_{k}p_{\textrm{cov},k}^{\textrm{MARP}}\left(\left\{ \lambda_{k}\right\} ,\left\{ T_{k}\right\} ,\left\{ B_{k}^{\textrm{U}}\right\} \right)\log_{2}\left(1+T_{k}\right).
\end{equation}
While the PT with the MIRP association scheme is similar except for
replacing $p_{\textrm{cov},k}^{\textrm{MARP}}$ by $p_{\textrm{cov},k}^{\textrm{MIRP}}$.

The EE can be derived by using Eq. (\ref{eq:EE}) and we will only
provide expressions for it when necessary.

\section{\label{sec:Performance Optimization}Performance Optimization and
Tradeoff}

As mentioned, from the mobile operators' point of view, the commercial
viability of network \lyxadded{Bean}{Thu Dec  7 10:09:24 2017}{densification}
depends on the underlying capital and operational expenditure \cite{Humar11Rethinking}.
While the former cost may be covered by taking up a high volume of
customers, with the rapid rise in the price of energy, and given that
BSs are particularly power-hungry, EE has become an increasingly crucial
factor for the success of dense HetNets \cite{Shojaeifard16Stochastic}.
\lyxadded{Bean}{Thu Dec  7 10:09:48 2017}{T}here are two main approaches
to enhance the energy consumption of cellular networks: 1) improvement
in hardware and 2) energy-efficient system design. The improvement
in hardware may have achieved its bottleneck due to the limit of Moore's
law, while the energy-efficient system design has a great potential
in the future 5G networks. In the following, two energy-efficient
optimization problems are proposed trying to obtain insights of the
system design.

\subsection{Optimizing coverage probability with the maximum total power consumption
constrain\lyxadded{Bean}{Thu Dec  7 05:51:05 2017}{t}}

\lyxadded{Bean}{Thu Dec  7 10:10:24 2017}{T}o pursue a further study
on coverage performance, we formulate a theoretical framework which
determines the optimal BS density to maximize the coverage probability
while guaranteeing that the total area power consumption is lower
than a given expected value $P^{\max}$ as follows
\begin{align}
\mathbf{OP1}:\underset{\lambda_{k}}{\max}\thinspace & p_{\textrm{cov}}^{\textrm{MARP}}\left(\left\{ \lambda_{k}\right\} ,\left\{ T_{k}\right\} ,\left\{ B_{k}^{\textrm{U}}\right\} \right)\nonumber \\
\textrm{s. t.}\thinspace & \textrm{C1: }\stackrel[k=1]{K}{\sum}\lambda_{k}\left(a_{k}P_{k}+b_{k}\right)\leqslant P^{\max}\label{eq:OP1}\\
\thinspace & \textrm{C2: }\lambda_{k}\geqslant0,\forall k\in\mathcal{K}\nonumber 
\end{align}
where $a_{k}$ and $b_{k}$ are defined in Eq. (\ref{eq:EE}). Note
that $\mathbf{OP1}$ assumes the MARP association scheme, while the
optimization problem with the MIRP association scheme is similar to
$\mathbf{OP1}$ and omitted here for brevity.

\subsection{Optimizing the EE under the minimum coverage probability constrain\lyxadded{Bean}{Thu Dec  7 05:51:09 2017}{t}}

In this subsection, another framework are formulated which determines
the optimal BS density to maximize the EE while guaranteeing QoS of
the network, \emph{i.e.}, the coverage probability is \lyxadded{Bean}{Thu Dec  7 10:11:39 2017}{higher}
than a given expected value $p_{\textrm{cov}}^{\min}$ as follows

\begin{align}
\mathbf{OP2}:\underset{\lambda_{k}}{\max}\thinspace & \mathcal{E}\left(\left\{ \lambda_{k}\right\} ,\left\{ T_{k}\right\} ,\left\{ B_{k}^{\textrm{U}}\right\} \right)\nonumber \\
\textrm{s. t.}\thinspace & \textrm{C1: }p_{\textrm{cov}}^{\textrm{MARP}}\left(\left\{ \lambda_{k}\right\} ,\left\{ T_{k}\right\} ,\left\{ B_{k}^{\textrm{U}}\right\} \right)\geqslant p_{\textrm{cov}}^{\min}\\
\thinspace & \textrm{C2: }\lambda_{k}\geqslant0,\forall k\in\mathcal{K}\nonumber 
\end{align}
We will show in the simulation results that tradeoff exists between
the coverage probability and the EE.

\subsection{Optimal deployment solution}

As NLoS and LoS transmissions are incorporated into our model, the
coverage probability is not a monotonically increasing function with
respect to BS density $\lambda_{k}$ like the cases in \cite{Cao13Optimal,Peng15Energy,Shojaeifard16Stochastic,Ge14Performance}
anymore. Besides, the coverage probability function is not convex
with respect to $\lambda_{k}$, either. Therefore, the optimization
problem under consideration should be tackled numerically. Exhaustive
search algorithms are well-suited for tackling the problem considering
that the objective function derivative is not available analytically
and its accurate evaluation is resource-intensive. Brent's algorithm
\cite{Brent73Algorithms} and heuristic downhill simplex method \cite{Press07Numerical}
can be utilized to obtain the solutions of $\mathbf{OP1}$ and $\mathbf{OP2}$
in exponential time. \lyxadded{Bean}{Thu Dec  7 10:12:53 2017}{T}o
gain an analytical insight into the effect of different operational
settings on the maximum energy-efficient deployment solution, in the
following\lyxadded{Bean}{Thu Dec  7 10:13:08 2017}{,} we focus on
the problem of finding the optimal BS density in a 2-tier HetNet.

\section{\label{sec:Results-and-Insights}Results and Insights}

A 2-tier HetNet is considered in our analysis. Macrocell BSs are in
Tier 1 and small cell BSs are in Tier 2. We assume that $P_{1}=46\textrm{ dBm}$,
$P_{2}=24\textrm{ dBm}$, $A_{1}^{\textrm{NL}}=2.7$, $A_{1}^{\textrm{L}}=30.8$,
$A_{2}^{\textrm{NL}}=32.9$, $A_{2}^{\textrm{L}}=41.1$, $\alpha_{1}^{\textrm{NL}}=4.28$,
$\alpha_{1}^{\textrm{L}}=2.42$, $\alpha_{2}^{\textrm{NL}}=3.75$,
$\alpha_{2}^{\textrm{L}}=2.09$, $\sigma_{1}^{\textrm{NL}}=8\textrm{ dB}$,
$\sigma_{1}^{\textrm{L}}=4\textrm{ dB}$, $\sigma_{2}^{\textrm{NL}}=4\textrm{ dB}$,
$\sigma_{2}^{\textrm{L}}=3\textrm{ dB}$, $\eta=-95\textrm{ dBm}$
\cite{Bai15Coverage,Singh15Tractable,3GPP36828,Yang17Density,Yang16Coverage,Ge17Multipath}
unless stated otherwise.

\subsection{Validation of the Analytical Results of Coverage Probability with
Monte Carlo Simulations}

\begin{figure}
\begin{centering}
\includegraphics[width=9cm]{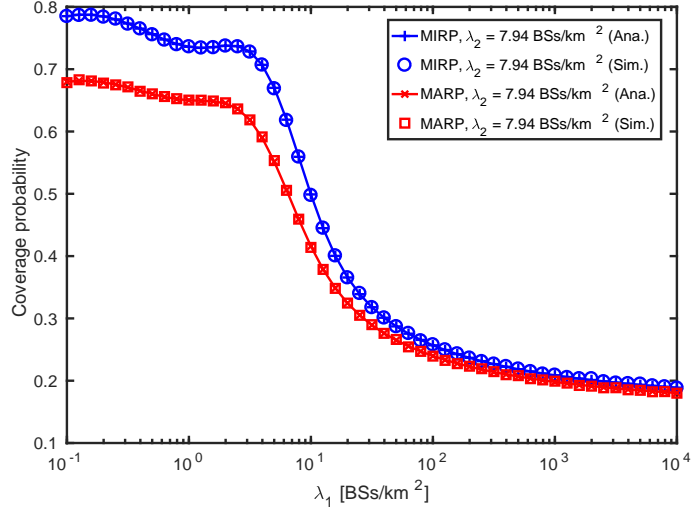}
\par\end{centering}
\caption{\label{fig:PC_ana_sim}Coverage probability vs. $\lambda_{1}$ with
the MIRP and MARP association schemes, $T_{k}=1\textrm{ dB}$.}
\end{figure}

\begin{figure}
\begin{centering}
\includegraphics[width=9cm]{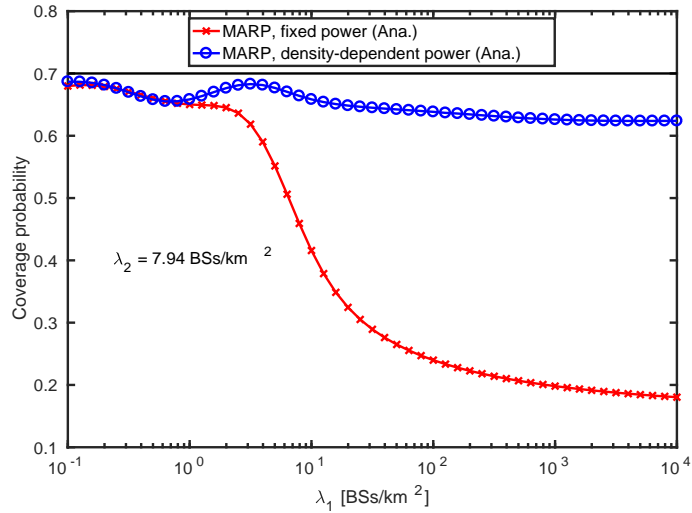}
\par\end{centering}
\caption{\label{fig:PC_fix_density}Coverage probability vs. $\lambda_{1}$
with the MARP association scheme, $T_{k}=1\textrm{ dB}$.}
\end{figure}

If fixing $\lambda_{2}$, the analytical and simulation results of
$p_{\textrm{cov}}^{\textrm{MIRP}}\left(\left\{ \lambda_{k}\right\} ,\left\{ T_{k}\right\} ,\left\{ B_{k}^{\textrm{U}}\right\} \right)$
and the analytical results of $p_{\textrm{cov}}^{\textrm{MARP}}\left(\left\{ \lambda_{k}\right\} ,\left\{ T_{k}\right\} ,\left\{ B_{k}^{\textrm{U}}\right\} \right)$
configured with $T=1\textrm{ dB}$ are plotted in Fig. \ref{fig:PC_ana_sim}
and Fig. \ref{fig:PC_fix_density}, respectively. As can be observed
from Fig. \ref{fig:PC_ana_sim}, the analytical results match the
simulation results well, which validate the accuracy of our theoretical
analysis. In Fig. \ref{fig:PC_fix_density}, aided by the utilization
of a density-dependent BS transmit power, the coverage probability
improves a lot as $\lambda_{1}$ increases.

\lyxadded{Bean}{Mon Nov 27 09:20:19 2017}{,,}

\begin{figure}
\begin{centering}
\includegraphics[width=9cm]{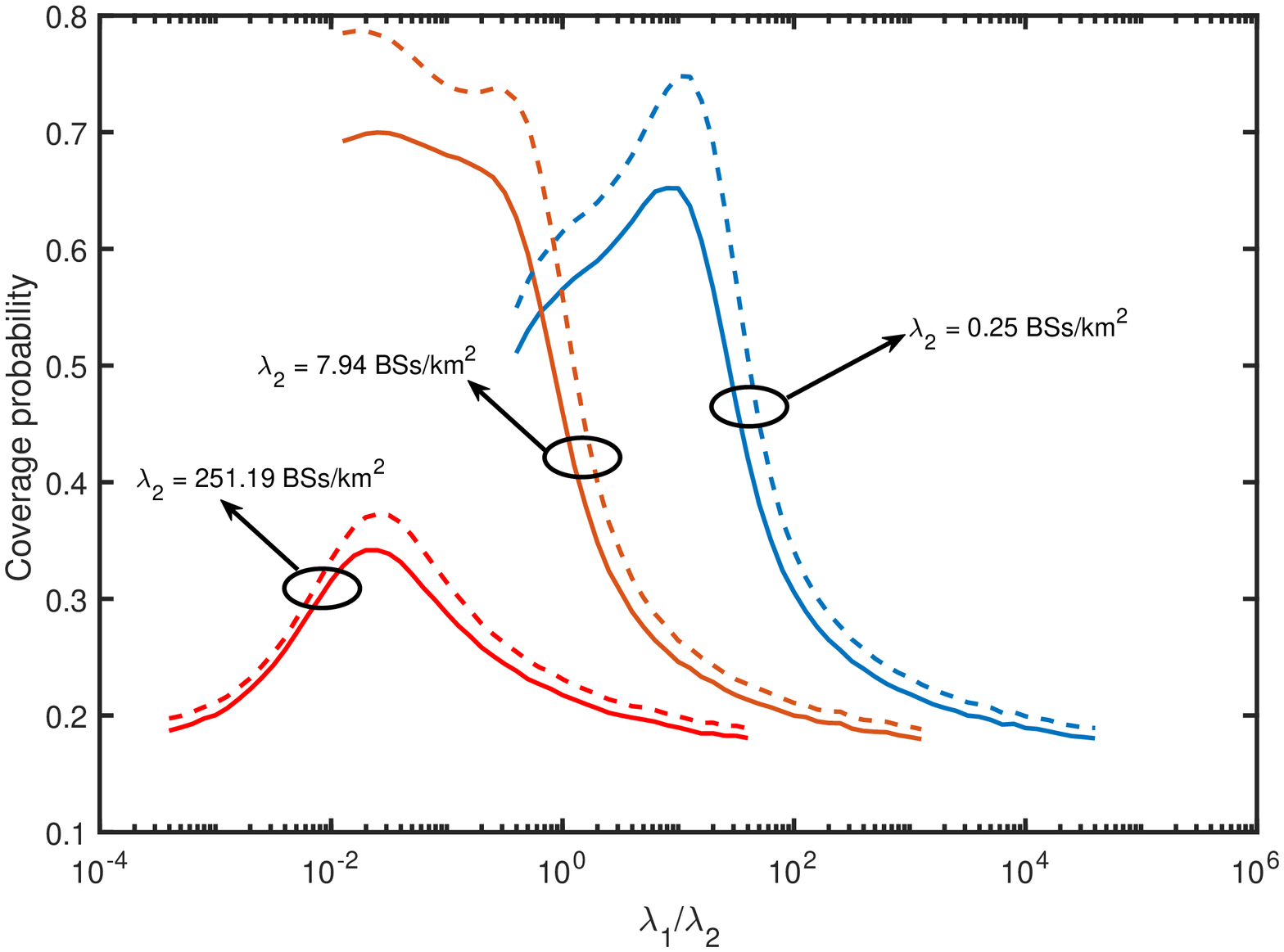}
\par\end{centering}
\caption{\label{fig:Pc_ratio_lambda1_fixed}Coverage probability vs. $\frac{\lambda_{1}}{\lambda_{2}}$
with the MARP \lyxadded{Bean}{Thu Dec  7 09:23:41 2017}{association
scheme (the solid line) and the MIRP association scheme (the dashed
line)} when $\lambda_{2}$ is fixed, $T_{k}=1\textrm{ dB}$.}
\end{figure}

\begin{figure}
\begin{centering}
\includegraphics[width=9cm]{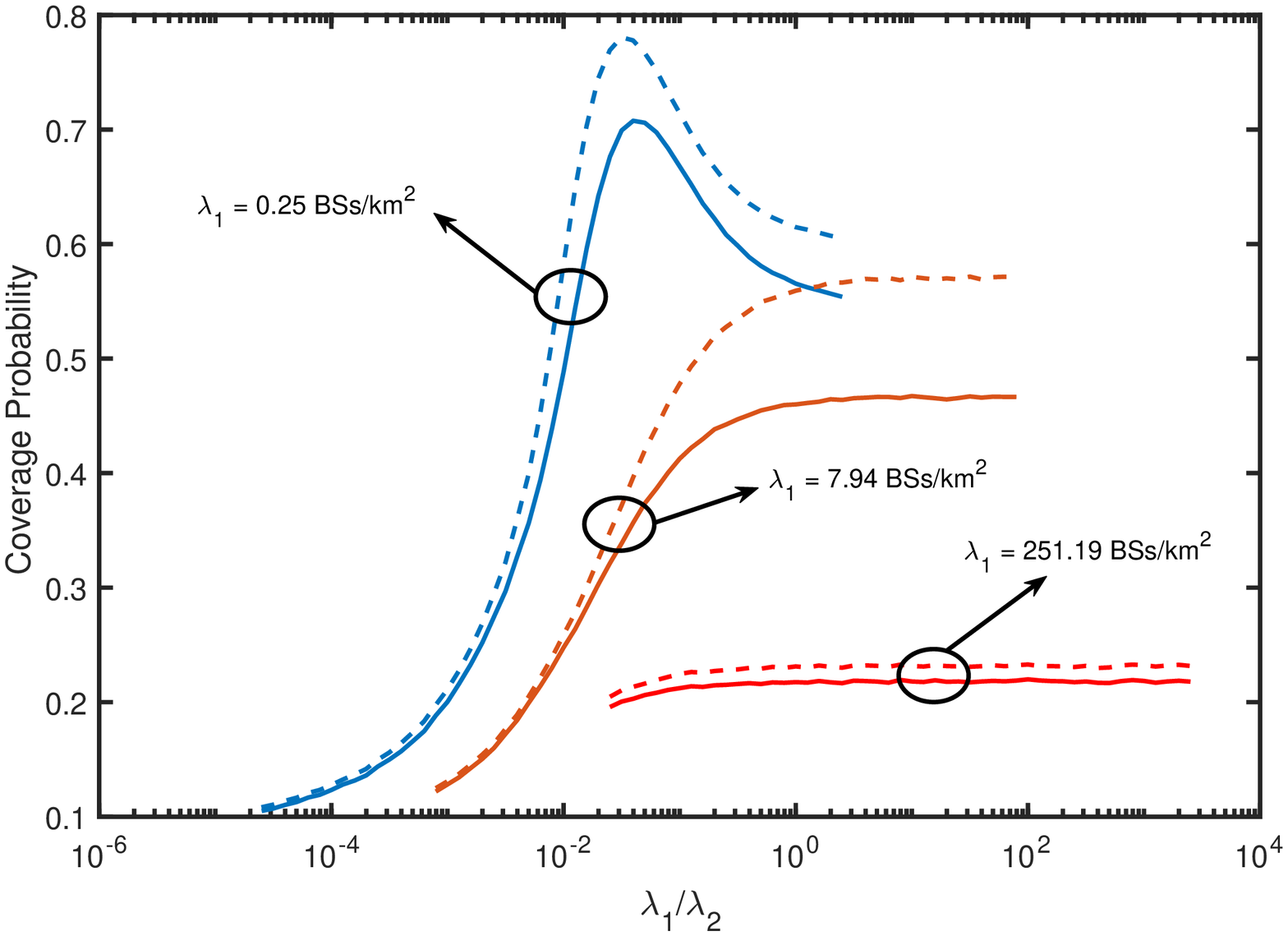}
\par\end{centering}
\caption{\label{fig:Pc_ratio_lambda2_fixed}Coverage probability vs. $\frac{\lambda_{1}}{\lambda_{2}}$
with the MARP \lyxadded{Bean}{Thu Dec  7 09:23:47 2017}{association
scheme (the solid line) and the MIRP association scheme (the dashed
line)} when $\lambda_{1}$ is fixed, $T_{k}=1\textrm{ dB}$.}
\end{figure}

Fig. \ref{fig:Pc_ratio_lambda1_fixed} and Fig. \ref{fig:Pc_ratio_lambda2_fixed}
illustrate the coverage probability vs. the ratio of $\lambda_{1}$
and $\lambda_{2}$, \emph{i.e.}, $\frac{\lambda_{1}}{\lambda_{2}}$
with the MIRP association scheme\lyxadded{Bean}{Sat Nov 25 07:11:21 2017}{
and the MIRP association scheme} when $\lambda_{1}$\lyxadded{Bean}{Sat Nov 25 07:11:51 2017}{
(or $\lambda_{2}$)} is fixed. It is found that in Fig. \ref{fig:Pc_ratio_lambda1_fixed},
there is always a coverage peak when $\frac{\lambda_{1}}{\lambda_{2}}$
is low, medium and high, \emph{i.e.}, $p_{\textrm{cov}}^{\max}=0.3417$\lyxadded{Bean}{Sun Nov 26 11:02:05 2017}{
(or 0.3725 with the MIRP)}, $p_{\textrm{cov}}^{\max}=0.6998$\lyxadded{Bean}{Sun Nov 26 11:02:35 2017}{
(or 0.7868 with the MIRP)}, $p_{\textrm{cov}}^{\max}=0.6521$\lyxadded{Bean}{Sun Nov 26 11:02:57 2017}{
(or 0.7476 with the MIRP)}, which indicates that there exists an optimal
$\lambda_{1}$ when implementing the network design if $\lambda_{2}$
is fixed. And in Fig. \ref{fig:Pc_ratio_lambda2_fixed}, the optimal
$\lambda_{2}$ exists as well. However, compared with Fig. \ref{fig:Pc_ratio_lambda1_fixed},
when \lyxadded{Bean}{Sun Nov 26 11:04:38 2017}{the fixed value of
}$\lambda_{1}$ is sparse, the coverage probability firstly increases
and then reaches a peak\lyxadded{Bean}{Thu Dec  7 10:14:58 2017}{.}
\lyxadded{Bean}{Thu Dec  7 10:15:04 2017}{F}inally it decreases to
a certain value. When\lyxadded{Bean}{Sun Nov 26 11:04:45 2017}{ the
fixed value of} $\lambda_{1}$ becomes larger, the coverage probability
saturates. Based on the above observations, dense deployment of small
cell BSs and macrocell BSs will lead to a better coverage probability.
However, there is no need to deploy an infinite number of BSs in a
finite area. When $\lambda_{1}$ approaches infinite\lyxadded{Bean}{Mon Nov 27 09:19:56 2017}{
if $\lambda_{2}$ remains fixed},\lyxadded{Bean}{Mon Nov 27 08:59:14 2017}{
and vice versa,} the coverage probability becomes much worse. In contrast,
\lyxadded{Bean}{Mon Nov 27 09:20:01 2017}{when $\lambda_{1}$ goes
to zero if $\lambda_{2}$ is fixed, and vice versa, the coverage probability
saturates to a certain value.}

\begin{figure}
\begin{centering}
\includegraphics[width=9cm]{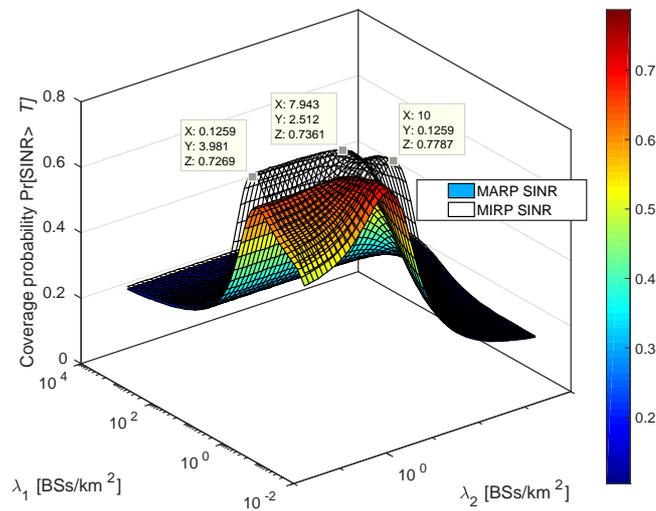}
\par\end{centering}
\caption{\label{fig:3D_Pc_MIRP_MARP}Comparison of coverage probability with
the MIRP and MARP association schemes.}
\end{figure}

\begin{figure}
\begin{centering}
\includegraphics[width=12cm]{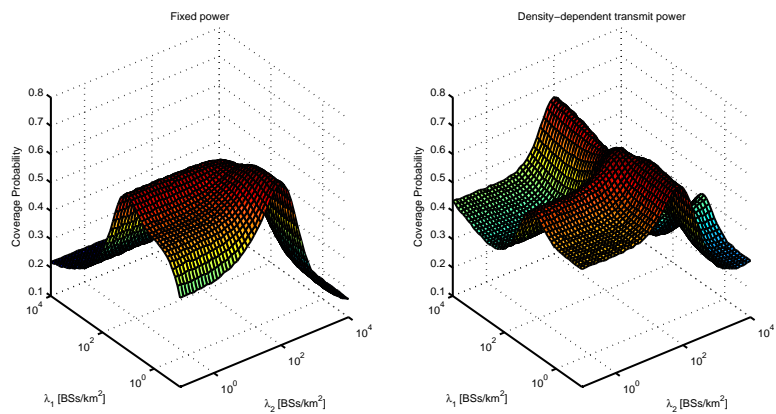}
\par\end{centering}
\caption{\label{fig:3D_Pc_fix_density}Comparison of coverage probability based
on different transmit power models, \emph{i.e.}, the fixed transmit
power and density-dependent transmit power.}
\end{figure}

To have a full picture of the coverage probability with respect to
$\lambda_{1}$ and $\lambda_{2}$, present two 3D figures are presented
in Fig. \ref{fig:3D_Pc_MIRP_MARP} and Fig. \ref{fig:3D_Pc_fix_density}.
In Fig. \ref{fig:3D_Pc_MIRP_MARP}, we compare the MIRP and MARP association
schemes based on the fixed transmit power. It is found that the coverage
probability with the MIRP association scheme is always greater than
that with the MARP association scheme as with former association scheme
BSs can provide the maximum power all the time even though it is not
practical in the real networks. In Fig. \ref{fig:3D_Pc_fix_density},
coverage probability based on the fixed transmit power and density-dependent
transmit power are illustrated, respectively. By utilizing a density-dependent
transmit power, the coverage probability improves compared with the
HetNets using \lyxadded{Bean}{Wed Nov 22 01:53:24 2017}{a} fixed transmit
power.\lyxadded{Bean}{Wed Nov 22 03:02:43 2017}{ Besides, it is noted
that the coverage probability using a density-dependent transmit power
fluctuates with BS density as illustrated in Fig. \ref{fig:3D_Pc_fix_density}
as well as in Fig. \ref{fig:PC_fix_density}. It is because the imperfect
power control used in Remark 1 which only depends on BS densities
and an approximate equivalent coverage area, the 3D coverage probability
appears more unique than that using a fixed transmit power.}

\subsection{The PT and the EE}

\begin{figure}
\begin{centering}
\includegraphics[width=12cm]{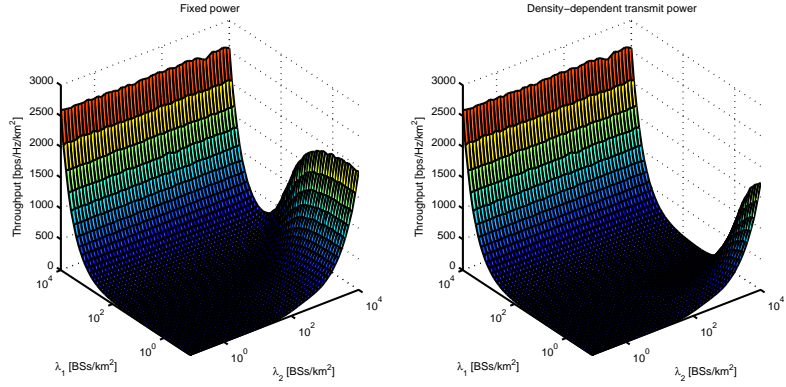}
\par\end{centering}
\caption{\label{fig:PT_density}The PT vs. $\lambda_{1}$ and $\lambda_{2}$
based on the fixed transmit power and density-dependent transmit power.}
\end{figure}

\begin{figure}
\begin{centering}
\includegraphics[width=12cm]{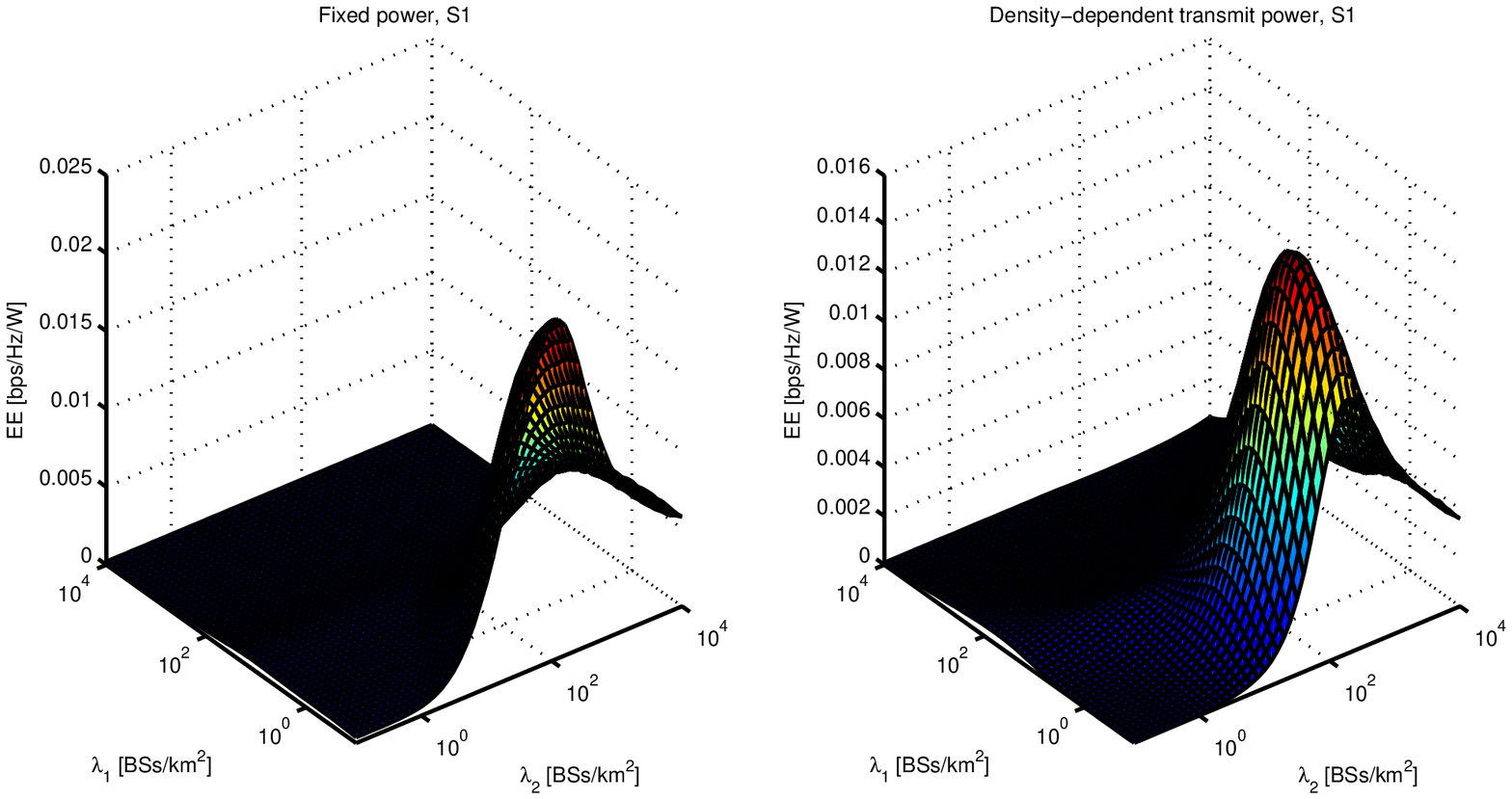}
\par\end{centering}
\caption{\label{fig:EE_density_S1}The EE vs. $\lambda_{1}$ and $\lambda_{2}$
based on the fixed transmit power and density-dependent transmit power
in scenario \textbf{S1}.}
\end{figure}

\begin{figure}
\begin{centering}
\includegraphics[width=12cm]{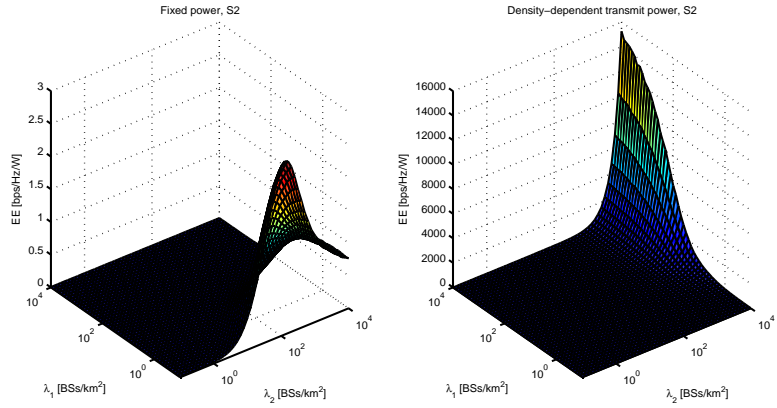}
\par\end{centering}
\caption{\label{fig:EE_density_S2}The EE vs. $\lambda_{1}$ and $\lambda_{2}$
based on the fixed transmit power and density-dependent transmit power
in scenario \textbf{S2}.}
\end{figure}

In this subsection, two typical energy consumption scenarios are considered,
\emph{i.e.}, practical power consumption and ideal power consumption,
denoted by \textbf{S1} and \textbf{S2}. Recall that the definition
of the EE in Eq. (\ref{eq:EE}) have parameters $\left\{ a_{k}\right\} $
and $\left\{ b_{k}\right\} $, thus we define \textbf{S1} as the HetNets
which are configured with $\left\{ a_{1}=22.6,a_{2}=5.5,b_{1}=414.2,b_{2}=32\right\} $
\cite{Peng15Energy} and \textbf{S2} configured with $\left\{ a_{1}=1,a_{2}=1,b_{1}=0,b_{2}=0\right\} $,
respectively. Note that \textbf{S2} accounts for the HetNets with
perfect power amplifier and ignoring the static power consumed by
signal processing, battery backup\lyxadded{Bean}{Thu Dec  7 10:16:11 2017}{,}
and cooling, \emph{etc}. In other words, in \textbf{S2} only radiated
power is considered. It is observed that $\lambda_{1}$ has a greater
impact on the PT than $\lambda_{2}$ in Fig. \ref{fig:PT_density}.
However, a larger $\lambda_{1}$ can not always provide a better EE
as illustrated in Fig. \ref{fig:EE_density_S1} and Fig. \ref{fig:EE_density_S2}.
Therefore, there should exist a tradeoff among coverage probability,
the PT and the EE, which is \lyxadded{Bean}{Thu Dec  7 10:16:39 2017}{revealed}
in the following subsection.

\subsection{Optimal Deployment Solutions}

\begin{figure}
\begin{centering}
\includegraphics[width=9cm]{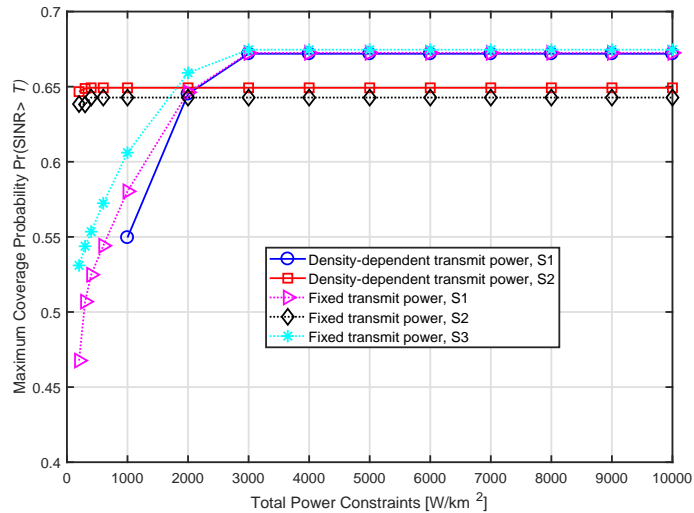}
\par\end{centering}
\caption{\label{fig:P1_PC} $\mathbf{OP1}$: The Maximum coverage probability
vs. $P^{\max}$ in scenarios \lyxadded{Bean}{Thu Dec  7 02:11:03 2017}{\textbf{S1},\textbf{
S2} and \textbf{S3} (\textbf{S3}: $\left\{ a_{1}=10.3,a_{2}=5.5,b_{1}=156.2,b_{2}=32\right\} $)}.}
\end{figure}

\begin{figure}
\begin{centering}
\includegraphics[width=9cm]{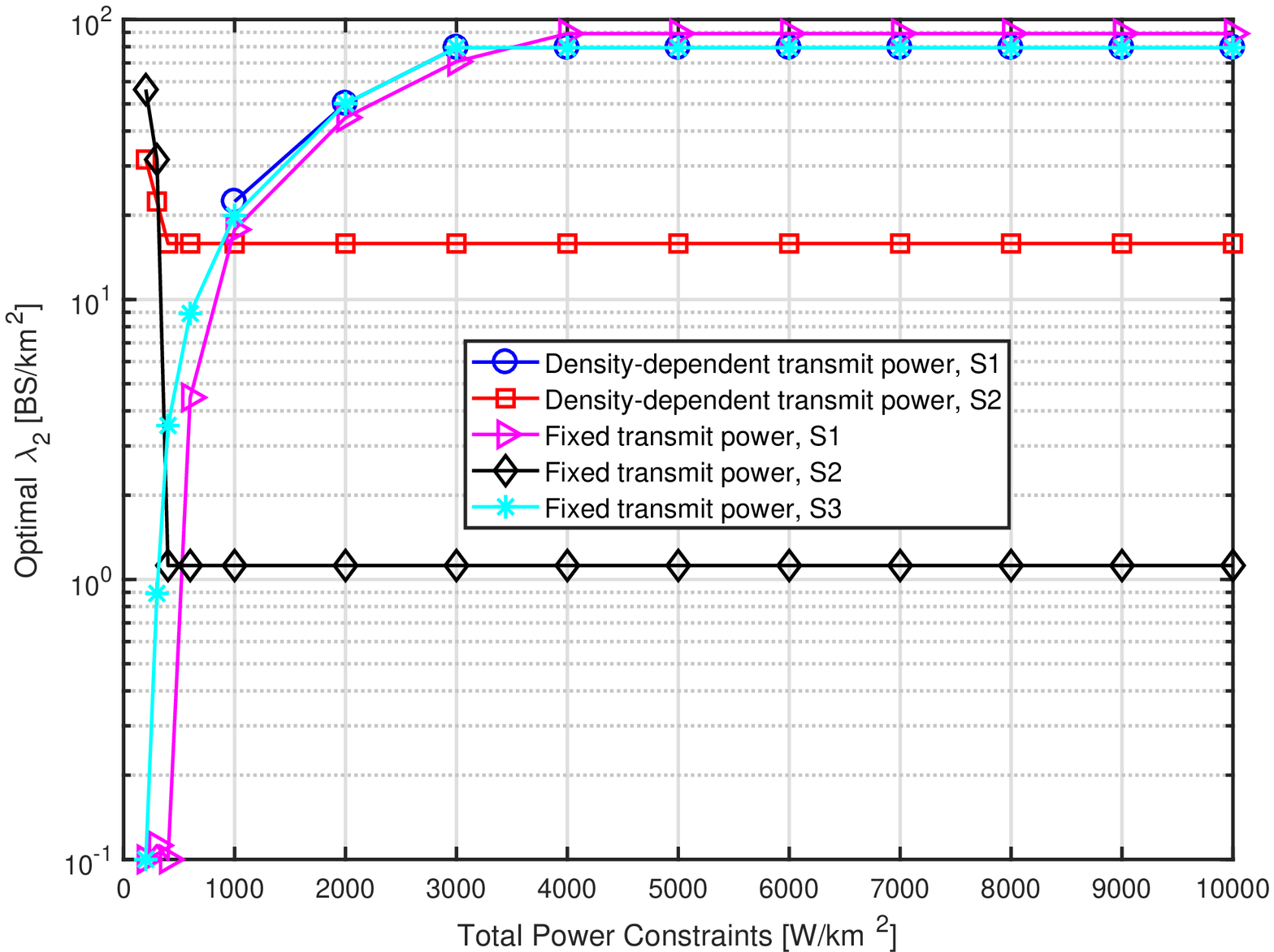}
\par\end{centering}
\caption{\label{fig:P1_lambda2}$\mathbf{OP1}$: The optimal $\lambda_{2}$
vs. $P^{\max}$ in scenarios \lyxadded{Bean}{Thu Dec  7 02:11:50 2017}{\textbf{S1},\textbf{
S2} and \textbf{S3} (\textbf{S3}: $\left\{ a_{1}=10.3,a_{2}=5.5,b_{1}=156.2,b_{2}=32\right\} $)}.}
\end{figure}

\begin{figure}
\begin{centering}
\includegraphics[width=9cm]{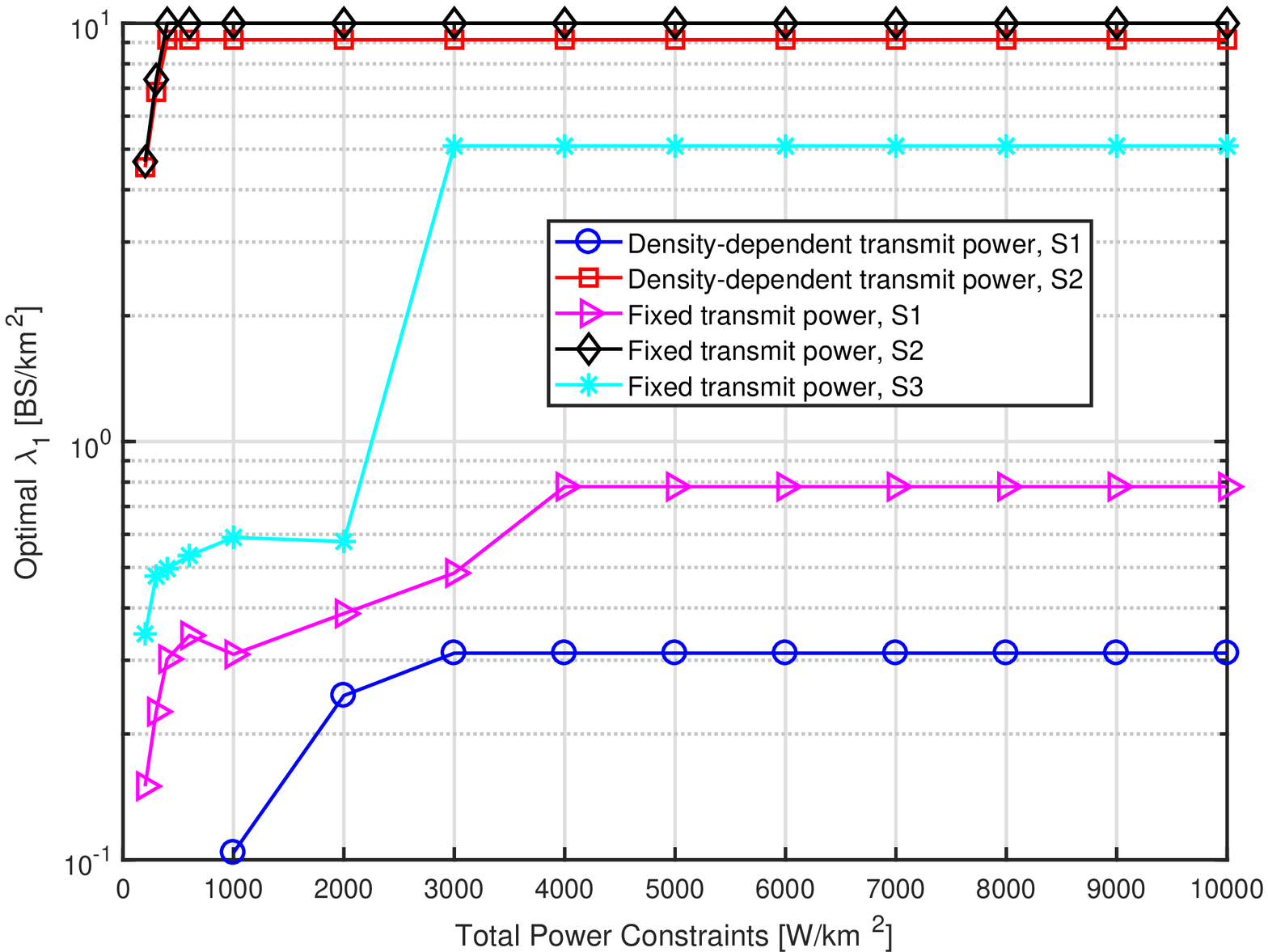}
\par\end{centering}
\caption{\label{fig:P1_lambda1}$\mathbf{OP1}$: The optimal $\lambda_{1}$
vs. $P^{\max}$ in scenarios \lyxadded{Bean}{Thu Dec  7 02:11:55 2017}{\textbf{S1},\textbf{
S2} and \textbf{S3} (\textbf{S3}: $\left\{ a_{1}=10.3,a_{2}=5.5,b_{1}=156.2,b_{2}=32\right\} $)}.}
\end{figure}

In this subsection, present the optimal deployment solutions for $\mathbf{OP1}$
and $\mathbf{OP2}$ are presented. Regarding $\mathbf{OP1}$, Fig.
\ref{fig:P1_PC}, Fig. \ref{fig:P1_lambda2} and Fig. \ref{fig:P1_lambda1}
offer the optimal coverage probability, the optimal $\lambda_{2}$
and the optimal $\lambda_{1}$ with respect to $P^{\max}$, respectively.
\lyxadded{Bean}{Fri Nov 17 08:48:54 2017}{From} Fig. \ref{fig:P1_PC},
we conclude that the maximum coverage probability increases with the
increase of $P^{\max}$ and finally becomes invariant with $P^{\max}$.
The reason behind this is that a larger $P^{\max}$ provides more
flexible BS deployment choice which will finally approach the optimal
BS deployment without the constrain\lyxadded{Bean}{Thu Dec  7 05:51:12 2017}{t}
of power consumption. Besides, the maximum coverage probability of
HetNets with a density-dependent transmit power is more sensitive
than that with a fixed transmit power. By comparison, the maximum
coverage probability in \textbf{S2} is superior to that in \textbf{S1}
when $P^{\max}$ is small and inferior to that in \textbf{S1} when
$P^{\max}$ becomes large. The optimal $\lambda_{2}$ in \textbf{S1}
grows up to a certain value with the increase of $P^{\max}$, after
which the optimal $\lambda_{2}$ reaches its saturation, as illustrated
in Fig. \ref{fig:P1_lambda2}. While in \textbf{S2}, the optimal $\lambda_{2}$
has an opposite tendency compared with that in \textbf{S1}. It is
because\lyxadded{Bean}{Thu Dec  7 10:19:12 2017}{,} in \textbf{S1},
static power consumption takes up most of the total power, especially
for the macrocell BSs. Therefore, deploying more small cell BSs can
save much more energy. If ignoring the static power consumption, \emph{i.e.},
\textbf{S2} is considered, every single macrocell BS can provide a
better coverage performance than every single small cell BS, thus
more macrocell BSs should be deployed in this scenario as shown in
Fig. \ref{fig:P1_lambda1}. 

\begin{figure}
\begin{centering}
\includegraphics[width=9cm]{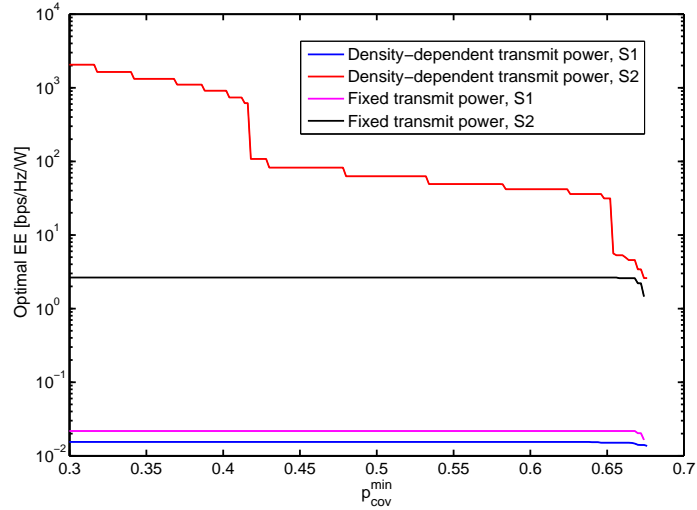}
\par\end{centering}
\caption{\label{fig:P2_EE}$\mathbf{OP2}$: The maximum EE vs. $p_{\textrm{cov}}^{\min}$
in scenarios \textbf{S1} and \textbf{S2}.}
\end{figure}

\begin{figure}
\begin{centering}
\includegraphics[width=9cm]{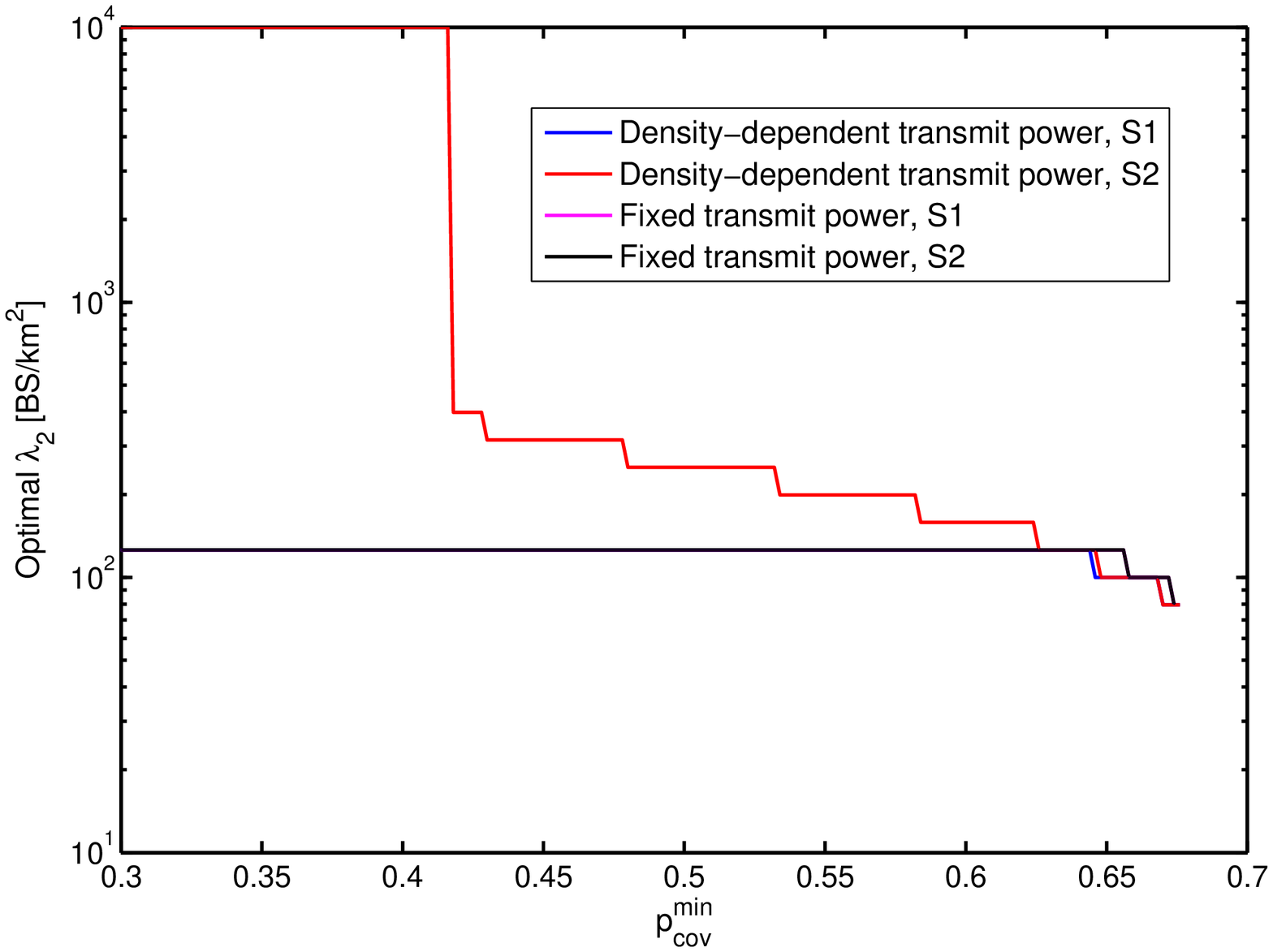}
\par\end{centering}
\caption{\label{fig:P2_lambda2}$\mathbf{OP2}$: The optimal $\lambda_{2}$
vs. $p_{\textrm{cov}}^{\min}$ in scenarios \textbf{S1} and \textbf{S2}.}
\end{figure}

\begin{figure}
\begin{centering}
\includegraphics[width=9cm]{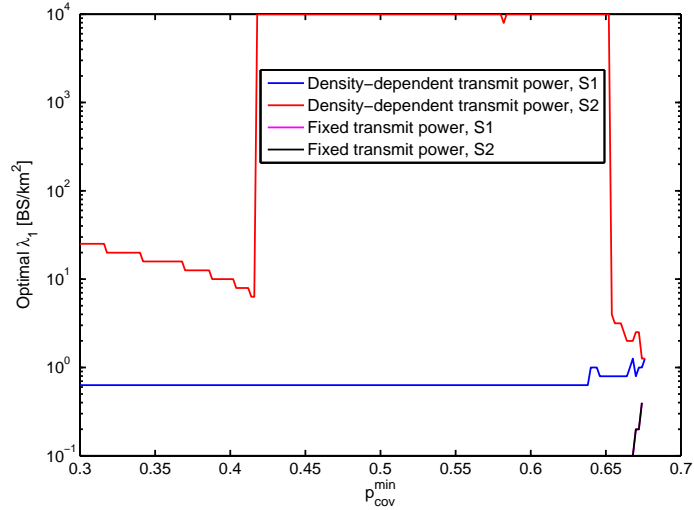}
\par\end{centering}
\caption{\label{fig:P2_lambda1}$\mathbf{OP2}$: The optimal $\lambda_{1}$
vs. $p_{\textrm{cov}}^{\min}$ in scenarios \textbf{S1} and \textbf{S2}.}
\end{figure}

Regarding $\mathbf{OP2}$, Fig. \ref{fig:P2_EE}, Fig. \ref{fig:P2_lambda2}
and Fig. \ref{fig:P2_lambda1} present the maximum EE, the optimal
$\lambda_{2}$ and $\lambda_{1}$, respectively. It is observed in
Fig. \ref{fig:P2_EE} that the maximum EE is a decreasing function
with respect to $p_{\textrm{cov}}^{\min}$. It is because a smaller
$p_{\textrm{cov}}^{\min}$ corresponds to a less constrain\lyxadded{Bean}{Thu Dec  7 05:49:57 2017}{t}
to the network deployment, as a result choosing proper BS densities
becomes much more feasible for mobile operators. The tendency of the
red curve, \emph{i.e.}, utilizing a density-dependent transmit power
in \textbf{S2}, is greatly different from the rest. It is because
the corresponding curve of the EE vs. $\lambda_{1}$ and $\lambda_{2}$
as illustrated in Fig. \ref{fig:EE_density_S1} and Fig. \ref{fig:EE_density_S2}
is different from that of the rest. \lyxadded{Bean}{Thu Dec  7 10:20:10 2017}{It
is also noted that the optimal EE is not strictly related to $p_{\textrm{cov}}^{\min}$,
\emph{e.g.}, when $0.43\leq p_{\textrm{cov}}^{\min}\leq0.48$ for
the red curve and when $0.30\leq p_{\textrm{cov}}^{\min}\leq0.656$
for the black curve, the optimal EE remains the same. It is because,
in these regimes, the optimal $\lambda_{1}$ and $\lambda_{2}$ can
guarantee the coverage probability is greater a bit more than the
threshold, \emph{i.e.}, $p_{\textrm{cov}}^{\min}$. To be specific,
when we deploy $\lambda_{1}=10^{4}\textrm{ BS/k\ensuremath{m^{2}}}$
and $\lambda_{2}=316.2\textrm{ BS/k\ensuremath{m^{2}}}$, the coverage
probability is 0.48. And if we set $p_{\textrm{cov}}^{\min}=0.43$,
the deployment of BSs, i.e., $\lambda_{1}=10^{4}\textrm{ BS/k\ensuremath{m^{2}}}$
and $\lambda_{2}=316.2\textrm{ BS/k\ensuremath{m^{2}}}$, can guarantee
the minimal coverage probability thus keeps the optimal EE the same.
}Moreover, when $p_{\textrm{cov}}^{\min}$ is greater than a certain
value, \emph{e.g.}, 0.6762 of the red curve, there is no feasible
solution to achieve the optimal EE as the QoS of the network, \emph{i.e.},
the coverage probability, can not be guaranteed. The optimal $\lambda_{2}$
is also a decreasing function with respect to $p_{\textrm{cov}}^{\min}$
in Fig. \ref{fig:P2_lambda2}. In Fig. \ref{fig:P2_lambda1}, when
$p_{\textrm{cov}}^{\min}$ is small, the network is not constrain\lyxadded{Bean}{Thu Dec  7 05:50:03 2017}{t}ed
by the coverage performance and deploying more small cell BSs can
achieve a better EE. While when $p_{\textrm{cov}}^{\min}$ is larger,
mobile operators have to deploy more macrocell BSs to guarantee the
network coverage performance, which results in a worse EE. The tendency
of the red curve in Fig. \ref{fig:P2_lambda1} is rather different
\lyxadded{Bean}{Wed Nov 29 08:16:58 2017}{from} others. When $p_{\textrm{cov}}^{\min}$
is small, the optimal $\lambda_{1}$ decreases with the increase of
$p_{\textrm{cov}}^{\min}$, then a ``flip-flop phenomenon'' appears,
\emph{i.e.}, the optimal $\lambda_{1}$ jumps to a high value to guarantee
the coverage performance and then decreases to a low value to achieve
high EE.\lyxadded{Bean}{Wed Nov 29 08:39:46 2017}{ Besides, comparing
Fig. \ref{fig:P2_lambda2} and Fig. \ref{fig:P2_lambda1}, it is found
that to achieve the optimal EE, an adjustment of $\lambda_{1}$ is
needed when $p_{\textrm{cov}}^{\min}$ is small, \emph{i.e.}, $0.30\leq p_{\textrm{cov}}^{\min}\leq0.42$,
while $\lambda_{2}$ may keep the same; an adjustment of $\lambda_{2}$
is needed when $p_{\textrm{cov}}^{\min}$ is medium, \emph{i.e.},
$0.42\leq p_{\textrm{cov}}^{\min}\leq0.65$, while $\lambda_{1}$
may keep the same; an adjustment of $\lambda_{1}$ as well as $\lambda_{2}$
is needed when $p_{\textrm{cov}}^{\min}$ is large, \emph{i.e.}, $p_{\textrm{cov}}^{\min}\geq0.65$.}

\section{\label{sec:Conclusions-and-Future}Conclusions and Future Work}

In this paper, we investigated network performance of downlink ultra-dense
HetNets and study the maximum energy-efficient BS deployment incorporating
both NLoS and LoS transmissions. Through analysis, we found that the
coverage probability with the MIRP association scheme is better than
that with the MARP association scheme and by utilizing a density-dependent
transmit power, the coverage probability improves when densities of
macrocell BSs and small cell BSs are sparse or medium compared with
the HetNets using the fixed transmit power. Moreover, we formulated
two optimization problems to achieve the maximum energy-efficient
deployment solution with certain minimum service criteria. Simulation
results show that there are tradeoffs among the coverage probability,
the total power consumption and the EE. In detail, the maximum coverage
probability with ideal power consumption is superior to that with
practical power consumption when the total power constrain\lyxadded{Bean}{Thu Dec  7 05:50:20 2017}{t}
is small and inferior to that with practical power consumption when
the total power constrain\lyxadded{Bean}{Thu Dec  7 05:50:24 2017}{t}
becomes large. Furthermore, the maximum EE is a decreasing function
with respect to the coverage probability constrain\lyxadded{Bean}{Thu Dec  7 05:50:29 2017}{t}.
In our future work, networks with idle mode capability and multiple-antennas
are also worth further studying.

\section*{Appendix A: Proof of Theorem 1}
\begin{IEEEproof}
The coverage probability in a $K$-tier HetNet with the MIRP association
scheme is defined as follows
\begin{equation}
p_{\textrm{cov}}^{\textrm{MIRP}}\left(\left\{ \lambda_{k}\right\} ,\left\{ T_{k}\right\} ,\left\{ B_{k}^{\textrm{U}}\right\} \right)=\Pr\Bigl[\underset{k\in\mathcal{K},\boldsymbol{X}_{k,o}\in\Phi_{k}}{\cup}\textrm{SINR}_{k}\left(\left\Vert \boldsymbol{X}_{k,o}\right\Vert \right)>T_{k}\Bigr].
\end{equation}
As we consider both NLoS and LoS transmissions, $p_{\textrm{cov}}^{\textrm{MIRP}}\left(\left\{ \lambda_{k}\right\} ,\left\{ T_{k}\right\} ,\left\{ B_{k}^{\textrm{U}}\right\} \right)$
can be further expressed by\lyxadded{Bean}{Thu Dec  7 10:38:44 2017}{
\begin{align*}
 & \quad\,p_{\textrm{cov}}^{\textrm{MIRP}}\left(\left\{ \lambda_{k}\right\} ,\left\{ T_{k}\right\} ,\left\{ B_{k}^{\textrm{U}}\right\} \right)=\Pr\Bigl[\underset{k\in\mathcal{K},\boldsymbol{X}_{k,o}\in\Phi_{k}}{\cup}\textrm{SINR}_{k}\left(\left\Vert \boldsymbol{X}_{k,o}\right\Vert \right)>T_{k}\Bigr]\\
 & =\mathbb{E}\left[\mathbb{I}\left(\underset{k\in\mathcal{K},\boldsymbol{X}_{k,o}\in\Phi_{k}}{\cup}\textrm{SINR}_{k}\left(\left\Vert \boldsymbol{X}_{k,o}\right\Vert \right)>T_{k}\right)\right]
\end{align*}
}
\begin{align}
 & =\underset{\textrm{I}}{\underbrace{\stackrel[k=1]{K}{\sum}\mathbb{E}\Bigl\{\underset{\boldsymbol{X}_{k,o}\in\Phi_{k}^{\textrm{NL}}}{\sum}\left[\mathbb{I}\left(\textrm{SINR}_{k}^{\textrm{NL}}\left(\left\Vert \boldsymbol{X}_{k,o}\right\Vert \right)>T_{k}\right)\right]\Bigr\}}}\nonumber \\
 & \quad\,+\underset{\textrm{II}}{\underbrace{\stackrel[k=1]{K}{\sum}\mathbb{E}\Bigl\{\underset{\boldsymbol{X}_{k,o}\in\Phi_{k}^{\textrm{L}}}{\sum}\left[\mathbb{I}\left(\textrm{SINR}_{k}^{\textrm{L}}\left(\left\Vert \boldsymbol{X}_{k,o}\right\Vert \right)>T_{k}\right)\right]\Bigr\}}},\label{eq:SINRNL}
\end{align}
where $\mathbb{I}\left(\cdot\right)$ is the indicator function, $\left(a\right)$
follows from \cite[Lemma 1]{Dhillon12Modeling} under the assumption
that $T_{k}\geqslant1\forall k$ and the independence between $\Phi_{k}^{\textrm{NL}}$
and $\Phi_{k}^{\textrm{L}}$, Part I and II in Eq. (\ref{eq:SINRNL})
can be comprehended as the probability that the typical MU is covered
by NLoS BSs and LoS BSs, respectively.

For Part I in Eq. (\ref{eq:SINRNL}), we have
\begin{align}
 & \quad\,\stackrel[k=1]{K}{\sum}\mathbb{E}\Bigl\{\underset{\boldsymbol{X}_{k,o}\in\Phi_{k}^{\textrm{NL}}}{\sum}\left[\mathbb{I}\left(\textrm{SINR}_{k}^{\textrm{NL}}\left(\left\Vert \boldsymbol{X}_{k,o}\right\Vert \right)>T_{k}\right)\right]\Bigr\}\overset{\left(a\right)}{=}\stackrel[k=1]{K}{\sum}\mathbb{E}\Bigl\{\underset{\overline{R_{k,o}^{\textrm{NL}}}\in\overline{\Phi_{k}^{\textrm{NL}}}}{\sum}\left[\mathbb{I}\left(\textrm{SINR}_{k}^{\textrm{NL}}\left(\overline{R_{k,o}^{\textrm{NL}}}\right)>T_{k}\right)\right]\Bigr\}\nonumber \\
 & \overset{\left(b\right)}{=}\stackrel[k=1]{K}{\sum}\int_{r=0}^{\infty}\Pr\Biggl[\frac{h_{k}^{\textrm{NL}}r^{-\alpha_{k}^{\textrm{NL}}}}{\stackrel[j=1]{K}{\sum}\underset{i:r_{j,i}\in\widehat{\Phi_{j}^{\textrm{NL}}}}{\sum}h_{j}^{\textrm{NL}}r_{j,i}^{-\alpha_{j}^{\textrm{NL}}}+\stackrel[j=1]{K}{\sum}\underset{i:r_{j,i}\in\overline{\Phi_{j}^{\textrm{L}}}}{\sum}h_{j}^{\textrm{L}}r_{j,i}^{-\alpha_{j}^{\textrm{L}}}+\eta}>T_{k}\Biggr]\lambda_{k}^{\textrm{NL}}\left(r\right)\textrm{d}r\nonumber \\
 & \overset{\left(c\right)}{=}\stackrel[k=1]{K}{\sum}\int_{r=0}^{\infty}\Pr\Biggl[\frac{h_{k}^{\textrm{NL}}r^{-\alpha_{k}^{\textrm{NL}}}}{\stackrel[j=1]{K}{\sum}I_{j}^{\textrm{NL}}+\stackrel[j=1]{K}{\sum}I_{j}^{\textrm{L}}+\eta}>T_{k}\Biggr]\lambda_{k}^{\textrm{NL}}\left(r\right)\textrm{d}r\nonumber \\
 & \overset{\left(d\right)}{=}\stackrel[k=1]{K}{\sum}\int_{r=0}^{\infty}e^{-T_{k}\eta r^{\alpha_{k}^{\textrm{NL}}}}\lambda_{k}^{\textrm{NL}}\left(r\right)\stackrel[j=1]{K}{\prod}\left[\mathcal{L}_{I_{j}^{\textrm{NL}}}^{\textrm{MIRP}}\left(T_{k}r^{\alpha_{k}^{\textrm{NL}}}\right)\mathcal{L}_{I_{j}^{\textrm{L}}}^{\textrm{MIRP}}\left(T_{k}r^{\alpha_{k}^{\textrm{NL}}}\right)\right]\textrm{d}r,
\end{align}
where $\left(a\right)$ is due to the transformation from $\Phi_{k}^{\textrm{NL}}$
to $\overline{\Phi_{k}^{\textrm{NL}}}$, $\left(b\right)$ follows
from Campbell theorem \cite{Haenggi12StochasticB} and variable substitution,
\emph{i.e.}, $\overline{R_{k,o}^{\textrm{NL}}}\rightarrow r$, in
$\left(c\right)$ $I_{j}^{\textrm{NL}}\overset{\textrm{def}}{=}\underset{i:r_{j,i}\in\overline{\Phi_{j}^{\textrm{NL}}}'}{\sum}h_{i}^{\textrm{NL}}r_{j,i}^{-\alpha_{j}^{\textrm{NL}}}$
and $I_{j}^{\textrm{L}}\overset{\textrm{def}}{=}\underset{i:r_{j,i}\in\overline{\Phi_{j}^{\textrm{NL}}}}{\sum}h_{i}^{\textrm{L}}r_{j,i}^{-\alpha_{j}^{\textrm{L}}}$
are the aggregate interference from NLoS BSs and LoS BSs in the $j$-th
tier, respectively, where $\widehat{\Phi_{j}^{\textrm{NL}}}=\overline{\Phi_{j}^{\textrm{NL}}}\setminus\left(0,\overline{R_{k,o}^{\textrm{NL}}}\right]$,
$\left(d\right)$ is due to $h_{k}^{\textrm{NL}}\sim\exp\left(1\right)$,
$\mathcal{L}_{I_{j}^{\textrm{NL}}}^{\textrm{MIRP}}\left(s\right)$
and $\mathcal{L}_{I_{j}^{\textrm{L}}}^{\textrm{MIRP}}\left(s\right)$
denote the Laplace transform of $I_{j}^{\textrm{NL}}$ and $I_{j}^{\textrm{L}}$
evaluated at $s$ with the MIRP association scheme, respectively.
Using the definition of Laplace transform, we derive $\mathcal{L}_{I_{j}^{\textrm{NL}}}^{\textrm{MIRP}}\left(s\right)$
as follows
\begin{align}
 & \quad\,\mathcal{L}_{I_{j}^{\textrm{NL}}}^{\textrm{MIRP}}\left(s\right)=\mathbb{E}_{I_{j}^{\textrm{NL}}}\left[e^{-sI_{j}^{\textrm{NL}}}\right]\overset{\left(a\right)}{=}\mathbb{E}_{\overline{\Phi_{j}^{\textrm{NL}}}}\biggl[\underset{i:r_{j,i}\in\widehat{\Phi_{j}^{\textrm{NL}}}}{\prod}\mathbb{E}_{h^{\textrm{NL}}}\left(e^{-sh^{\textrm{NL}}r_{j,i}^{-\alpha_{j}^{\textrm{NL}}}}\right)\biggr]\nonumber \\
 & =\mathbb{E}_{\overline{\Phi_{j}^{\textrm{NL}}}}\biggl[\underset{i:r_{j,i}\in\widehat{\Phi_{j}^{\textrm{NL}}}}{\prod}\frac{1}{1+sr_{j,i}^{-\alpha_{j}^{\textrm{NL}}}}\biggr]\overset{\left(b\right)}{=}\exp\biggl[\int_{y=0}^{\infty}\left(\frac{1}{1+sy^{-\alpha_{j}^{\textrm{NL}}}}-1\right)\lambda_{j}^{\textrm{NL}}\left(y\right)\textrm{d}y\biggr]\nonumber \\
 & =\exp\biggl[-\int_{y=0}^{\infty}\frac{\lambda_{j}^{\textrm{NL}}\left(y\right)}{1+y^{\alpha_{j}^{\textrm{NL}}}/s}\textrm{d}y\biggr]=\exp\biggl[-s^{1/\alpha_{j}^{\textrm{NL}}}\int_{y=0}^{\infty}\frac{\lambda_{j}^{\textrm{NL}}\left(ys^{1/\alpha_{j}^{\textrm{NL}}}\right)}{1+y^{\alpha_{j}^{\textrm{NL}}}}\textrm{d}y\biggr],\label{eq:Proof_LT_NL_max_ins}
\end{align}
where $\left(a\right)$ follows from the independence between the
fading random variables (RVs), \emph{i.e.}, $h_{j}^{\textrm{NL}}$,
$\left(b\right)$ follows from probability generating functional (PGFL)
of PPP \cite{Haenggi12StochasticB}.

Similarly, $\mathcal{L}_{I_{j}^{\textrm{L}}}^{\textrm{MIRP}}\left(s\right)$
is obtained as follows
\begin{align}
 & \quad\,\mathcal{L}_{I_{j}^{\textrm{L}}}^{\textrm{MIRP}}\left(s\right)=\mathbb{E}_{I_{j}^{\textrm{L}}}\left[e^{-sI_{j}^{\textrm{L}}}\right]=\exp\biggl[-\int_{y=0}^{\infty}\frac{\lambda_{j}^{\textrm{L}}\left(y\right)}{1+y^{\alpha_{j}^{\textrm{L}}}/s}\textrm{d}y\biggr]\nonumber \\
 & =\exp\biggl[-s^{1/\alpha_{j}^{\textrm{L}}}\int_{y=0}^{\infty}\frac{\lambda_{j}^{\textrm{L}}\left(ys^{1/\alpha_{j}^{\textrm{L}}}\right)}{1+y^{\alpha_{j}^{\textrm{L}}}}\textrm{d}y\biggr],\label{eq:Proof_LT_L_max_ins}
\end{align}

Using a similar approach compared with Part I, Part II can also be
easily obtained as follows
\begin{align}
 & \quad\,\stackrel[k=1]{K}{\sum}\mathbb{E}\left\{ \underset{\boldsymbol{X}_{k,o}\in\Phi_{k}^{\textrm{L}}}{\sum}\left[\mathbb{I}\left(\textrm{SINR}^{\textrm{L}}\left(\left\Vert \boldsymbol{X}_{k,o}\right\Vert \right)>T_{k}\right)\right]\right\} \nonumber \\
 & =\stackrel[k=1]{K}{\sum}\int_{t=0}^{\infty}e^{-T_{k}\eta r^{\alpha_{k}^{\textrm{L}}}}\lambda_{k}^{\textrm{L}}\left(r\right)\stackrel[j=1]{K}{\prod}\left[\mathcal{L}_{I_{j}^{\textrm{NL}}}^{\textrm{MIRP}}\left(T_{k}r^{\alpha_{k}^{\textrm{L}}}\right)\mathcal{L}_{I_{j}^{\textrm{L}}}^{\textrm{MIRP}}\left(T_{k}r^{\alpha_{k}^{\textrm{L}}}\right)\right]\textrm{d}r,
\end{align}
where $\mathcal{L}_{I_{j}^{\textrm{NL}}}^{\textrm{MIRP}}\left(s\right)$
and $\mathcal{L}_{I_{j}^{\textrm{L}}}^{\textrm{MIRP}}\left(s\right)$
are defined in Eq. (\ref{eq:Proof_LT_NL_max_ins}) and Eq. (\ref{eq:Proof_LT_L_max_ins}),
respectively. Then, the proof is completed.
\end{IEEEproof}

\section*{Appendix B: Proof of Theorem 2}

Using the law of total probability, we can calculate coverage probability
$p_{\textrm{cov}}^{\textrm{MARP}}\left(\left\{ \lambda_{k}\right\} ,\left\{ T_{k}\right\} ,\left\{ B_{k}^{\textrm{U}}\right\} \right)$
as 
\begin{equation}
p_{\textrm{cov}}^{\textrm{MARP}}\left(\left\{ \lambda_{k}\right\} ,\left\{ T_{k}\right\} ,\left\{ B_{k}^{\textrm{U}}\right\} \right)=\stackrel[k=1]{K}{\sum}p_{k}^{\textrm{NL}}\left(\left\{ \lambda_{k}^{\textrm{NL}}\right\} ,\left\{ T_{k}\right\} ,\left\{ B_{k}^{\textrm{NL}}\right\} \right)+\stackrel[k=1]{K}{\sum}p_{k}^{\textrm{L}}\left(\left\{ \lambda_{k}^{\textrm{L}}\right\} ,\left\{ T_{k}\right\} ,\left\{ B_{k}^{\textrm{L}}\right\} \right),
\end{equation}
where the first part and the second part on the right side of the
equation denote the conditional coverage probability that the typical
MU is in the coverage of NLoS BSs and LoS\lyxadded{Bean}{Fri Nov 17 07:28:40 2017}{
BS}s, respectively, by observing that the two events are disjoint.
Given that the typical MU is served by a\lyxadded{Bean}{Thu Dec  7 09:21:44 2017}{n}
NLoS BS and an the maximum average received power is denote by $\mathcal{P}_{k}^{\textrm{NL}}$,
\emph{i.e.}, $\mathcal{P}_{k}^{\textrm{NL}}=\max\left(P_{k.i}^{\textrm{NL}}\right)$.
Then 
\begin{align}
 & \quad\,p_{k}^{\textrm{NL}}\left(\left\{ \lambda_{k}^{\textrm{NL}}\right\} ,\left\{ T_{k}\right\} ,\left\{ B_{k}^{\textrm{NL}}\right\} \right)\nonumber \\
 & =\Pr\biggl[\left(\textrm{SINR}_{k}^{\textrm{NL}}>T_{k}\right)\cap\left(\underset{j\in\mathcal{K}}{\cap}\left\{ \mathcal{P}_{k}^{\textrm{NL}}>\mathcal{P}_{j}^{\textrm{L}}\right\} ,\underset{j\in\mathcal{K}\setminus k}{\cap}\left\{ \mathcal{P}_{k}^{\textrm{NL}}>\mathcal{P}_{j}^{\textrm{NL}}\right\} \right)\cap\mathcal{Y}_{k}^{\textrm{NL}}\biggr]\nonumber \\
 & =\mathbb{E}_{\mathcal{Y}_{k}^{\textrm{NL}}}\biggl\{\underset{\textrm{II}}{\underbrace{\Pr\biggl[\textrm{SINR}_{k}^{\textrm{NL}}>T_{k}\left|\left(\underset{j\in\mathcal{K}}{\cap}\left\{ \mathcal{P}_{k}^{\textrm{NL}}>\mathcal{P}_{j}^{\textrm{L}}\right\} ,\underset{j\in\mathcal{K}\setminus k}{\cap}\left\{ \mathcal{P}_{k}^{\textrm{NL}}>\mathcal{P}_{j}^{\textrm{NL}}\right\} \right)\cap\mathcal{Y}_{k}^{\textrm{NL}}\right.\biggr]}}\nonumber \\
 & \quad\,\times\underset{\textrm{I}}{\underbrace{\Pr\biggl[\left.\underset{j\in\mathcal{K}}{\cap}\left\{ \mathcal{P}_{k}^{\textrm{NL}}>\mathcal{P}_{j}^{\textrm{L}}\right\} ,\underset{j\in\mathcal{K}\setminus k}{\cap}\left\{ \mathcal{P}_{k}^{\textrm{NL}}>\mathcal{P}_{j}^{\textrm{NL}}\right\} \right|\mathcal{Y}_{k}^{\textrm{NL}}\biggr]}}\biggr\},\label{eq:proof_pcNL}
\end{align}
where $\mathcal{Y}_{k}^{\textrm{NL}}$ is the equivalent distance
between the typical MU and the BS providing the maximum average received
power to the typical MU in $\Phi_{k}^{\textrm{NL}}$, \emph{i.e.},
$\mathcal{Y}_{k}^{\textrm{NL}}=\underset{\overline{R_{k.i}^{\textrm{NL}}}\in\overline{\Phi_{k}^{\textrm{NL}}}}{\arg\max}\left(\overline{R_{k.i}^{\textrm{NL}}}\right)^{-\alpha_{k}^{\textrm{NL}}}$,
and also note that $\mathcal{P}_{k}^{\textrm{NL}}=\left(\mathcal{Y}_{k}^{\textrm{NL}}\right)^{-\alpha_{k}^{\textrm{NL}}}$.
For Part I, 
\begin{align}
 & \quad\,\Pr\left[\left.\underset{j\in\mathcal{K}}{\cap}\left\{ \mathcal{P}_{k}^{\textrm{NL}}>\mathcal{P}_{j}^{\textrm{L}}\right\} ,\underset{j\in\mathcal{K}\setminus k}{\cap}\left\{ \mathcal{P}_{k}^{\textrm{NL}}>\mathcal{P}_{j}^{\textrm{NL}}\right\} \right|\mathcal{Y}_{k}^{\textrm{NL}}\right]\nonumber \\
 & =\underset{j\in\mathcal{K}}{\prod}\Pr\left[\left.\mathcal{P}_{k}^{\textrm{NL}}>\mathcal{P}_{j}^{\textrm{L}}\right|\mathcal{Y}_{k}^{\textrm{NL}}\right]\underset{j\in\mathcal{K}\setminus k}{\prod}\Pr\left[\left.\mathcal{P}_{k}^{\textrm{NL}}>\mathcal{P}_{j}^{\textrm{NL}}\right|\mathcal{Y}_{k}^{\textrm{NL}}\right]\nonumber \\
 & =\underset{j\in\mathcal{K}}{\prod}\Pr\left[\left.\left(\mathcal{Y}_{k}^{\textrm{NL}}\right)^{-\alpha_{k}^{\textrm{NL}}}>\left(\mathcal{Y}_{j}^{\textrm{L}}\right)^{-\alpha_{j}^{\textrm{L}}}\right|\mathcal{Y}_{k}^{\textrm{NL}}\right]\underset{j\in\mathcal{K}\setminus k}{\prod}\Pr\left[\left.\left(\mathcal{Y}_{k}^{\textrm{NL}}\right)^{-\alpha_{k}^{\textrm{NL}}}>\left(\mathcal{Y}_{j}^{\textrm{NL}}\right)^{-\alpha_{j}^{\textrm{NL}}}\right|\mathcal{Y}_{k}^{\textrm{NL}}\right]\nonumber \\
 & =\underset{j\in\mathcal{K}}{\prod}\Pr\left[\left.\mathcal{Y}_{j}^{\textrm{L}}>\left(\mathcal{Y}_{k}^{\textrm{NL}}\right)^{\alpha_{k}^{\textrm{NL}}/\alpha_{j}^{\textrm{L}}}\right|\mathcal{Y}_{k}^{\textrm{NL}}\right]\underset{j\in\mathcal{K}\setminus k}{\prod}\Pr\left[\left.\mathcal{Y}_{j}^{\textrm{NL}}>\left(\mathcal{Y}_{k}^{\textrm{NL}}\right)^{\alpha_{k}^{\textrm{NL}}/\alpha_{j}^{\textrm{NL}}}\right|\mathcal{Y}_{k}^{\textrm{NL}}\right]\nonumber \\
 & \overset{\left(a\right)}{=}\underset{j\in\mathcal{K}}{\prod}\exp\left[-\Lambda_{j}^{\textrm{L}}\left(\left[0,\left(\mathcal{Y}_{k}^{\textrm{NL}}\right)^{\alpha_{k}^{\textrm{NL}}/\alpha_{j}^{\textrm{L}}}\right]\right)\right]\underset{j\in\mathcal{K}\setminus k}{\prod}\exp\left[-\Lambda_{j}^{\textrm{NL}}\left(\left[0,\left(\mathcal{Y}_{k}^{\textrm{NL}}\right)^{\alpha_{k}^{\textrm{NL}}/\alpha_{j}^{\textrm{NL}}}\right]\right)\right]\nonumber \\
 & =\exp\biggl\{-\underset{j\in\mathcal{K}}{\sum}\left[\Lambda_{j}^{\textrm{L}}\left(\left[0,\left(\mathcal{Y}_{k}^{\textrm{NL}}\right)^{\alpha_{k}^{\textrm{NL}}/\alpha_{j}^{\textrm{L}}}\right]\right)+\Lambda_{j}^{\textrm{NL}}\left(\left[0,\left(\mathcal{Y}_{k}^{\textrm{NL}}\right)^{\alpha_{k}^{\textrm{NL}}/\alpha_{j}^{\textrm{NL}}}\right]\right)\right]+\Lambda_{k}^{\textrm{NL}}\left(\left[0,\mathcal{Y}_{k}^{\textrm{NL}}\right]\right)\biggr\},\label{eq:proof_PNL g PL}
\end{align}
where $\mathcal{Y}_{k}^{\textrm{L}}$, similar to the definition of
$\mathcal{Y}_{k}^{\textrm{NL}}$, is the equivalent distance between
the typical MU and the BS providing the maximum average received power
to the typical MU in $\Phi_{k}^{\textrm{L}}$, \emph{i.e.}, $\mathcal{Y}_{k}^{\textrm{L}}=\underset{\overline{R_{k,i}^{\textrm{L}}}\in\overline{\Phi_{k}^{\textrm{L}}}}{\arg\max}\left(\overline{R_{k,i}^{\textrm{L}}}\right)^{-\alpha_{k}^{\textrm{L}}}$,
and also note that $\mathcal{P}_{k}^{\textrm{L}}=\left(\mathcal{Y}_{k}^{\textrm{L}}\right)^{-\alpha_{k}^{\textrm{L}}}$,
and $\left(a\right)$ follows from the void probability of a PPP.

For Part II, we know that $\textrm{SINR}_{k}^{\textrm{NL}}=\frac{h_{k}^{\textrm{NL}}\mathcal{P}_{k}^{\textrm{NL}}}{\stackrel[j=1]{K}{\sum}I_{j}^{\textrm{NL}}+\stackrel[j=1]{K}{\sum}I_{j}^{\textrm{L}}+\eta}$.
The conditional coverage probability can be derived as follows
\begin{align}
 & \quad\,\Pr\biggl[\textrm{SINR}_{k}^{\textrm{NL}}>T_{k}\left|\left(\underset{j\in\mathcal{K}}{\cap}\left\{ \mathcal{P}_{k}^{\textrm{NL}}>\mathcal{P}_{j}^{\textrm{L}}\right\} ,\underset{j\in\mathcal{K}\setminus k}{\cap}\left\{ \mathcal{P}_{k}^{\textrm{NL}}>\mathcal{P}_{j}^{\textrm{NL}}\right\} \right)\cap\mathcal{Y}_{k}^{\textrm{NL}}\right.\biggr]\nonumber \\
 & \overset{\left(a\right)}{=}\Pr\biggl[\frac{h_{k}^{\textrm{NL}}\left(\mathcal{Y}_{k}^{\textrm{NL}}\right)^{-\alpha_{k}^{\textrm{NL}}}}{\stackrel[j=1]{K}{\sum}I_{j}^{\textrm{NL}}+\stackrel[j=1]{K}{\sum}I_{j}^{\textrm{L}}+\eta}>T_{k}\biggl|\textrm{E}\biggr]\nonumber \\
 & =\Pr\biggl[h_{k}^{\textrm{NL}}>T_{k}\left(\mathcal{Y}_{k}^{\textrm{NL}}\right)^{\alpha_{k}^{\textrm{NL}}}\left(\stackrel[j=1]{K}{\sum}I_{j}^{\textrm{NL}}+\stackrel[j=1]{K}{\sum}I_{j}^{\textrm{L}}+\eta\right)\biggl|\textrm{E}\biggr]\nonumber \\
 & \overset{\left(b\right)}{=}e^{-T_{k}\eta r^{\alpha_{k}^{\textrm{NL}}}}\stackrel[j=1]{K}{\prod}\left[\mathcal{L}_{I_{j}^{\textrm{NL}}}^{\textrm{MARP1}}\left(T_{k}r^{\alpha_{k}^{\textrm{NL}}}\right)\mathcal{L}_{I_{j}^{\textrm{L}}}^{\textrm{MARP1}}\left(T_{k}r^{\alpha_{k}^{\textrm{NL}}}\right)\right],\label{eq:proof_SINR}
\end{align}
where in $\left(a\right)$ event $\textrm{E}\overset{\textrm{def}}{=}\left(\underset{j\in\mathcal{K}}{\cap}\left\{ \mathcal{P}_{k}^{\textrm{NL}}>\mathcal{P}_{j}^{\textrm{L}}\right\} ,\underset{j\in\mathcal{K}\setminus k}{\cap}\left\{ \mathcal{P}_{k}^{\textrm{NL}}>\mathcal{P}_{j}^{\textrm{NL}}\right\} \right)\cap\mathcal{Y}_{k}^{\textrm{NL}}$,
$\left(b\right)$ follows from $h_{k}^{\textrm{NL}}\sim\exp\left(1\right)$
and variable substitution, \emph{i.e.}, $\mathcal{Y}_{k}^{\textrm{NL}}\rightarrow r$,
$\mathcal{L}_{I_{j}^{\textrm{NL}}}^{\textrm{ave}}\left(s\right)$
and $\mathcal{L}_{I_{j}^{\textrm{L}}}^{\textrm{ave}}\left(s\right)$
denote the Laplace transform of $I_{j}^{\textrm{NL}}$ and $I_{j}^{\textrm{L}}$
evaluated at $s$ with the MARP association scheme, respectively.
Like Appendix A, we derive $\mathcal{L}_{I_{j}^{\textrm{NL}}}^{\textrm{MARP}}\left(s\right)$
as follows
\begin{align}
 & \quad\,\mathcal{L}_{I_{j}^{\textrm{NL}}}^{\textrm{MARP1}}\left(s\right)=\mathbb{E}_{I_{j}^{\textrm{NL}}}\left[e^{-sI_{j}^{\textrm{NL}}}\right]=\mathbb{E}_{\overline{\Phi_{j}^{\textrm{NL}}}}\biggl[\underset{i:r_{j,i}\in\widehat{\Phi_{j}^{\textrm{NL}}}}{\prod}\mathbb{E}_{h^{\textrm{NL}}}\left(e^{-sh^{\textrm{NL}}r_{j,i}^{-\alpha_{j}^{\textrm{NL}}}}\right)\biggr]\nonumber \\
 & =\mathbb{E}_{\overline{\Phi_{j}^{\textrm{NL}}}}\biggl[\underset{i:r_{j,i}\in\widehat{\Phi_{j}^{\textrm{NL}}}}{\prod}\frac{1}{1+sr_{j,i}^{-\alpha_{j}^{\textrm{NL}}}}\biggr]\overset{\left(a\right)}{=}\exp\biggl[\int_{y=\left(\mathcal{Y}_{k}^{\textrm{NL}}\right)^{\alpha_{k}^{\textrm{NL}}/\alpha_{j}^{\textrm{NL}}}}^{\infty}\left(\frac{1}{1+sy^{-\alpha_{j}^{\textrm{NL}}}}-1\right)\lambda_{j}^{\textrm{NL}}\left(y\right)\textrm{d}y\biggr]\nonumber \\
 & =\exp\biggl[-\int_{y=\left(\mathcal{Y}_{k}^{\textrm{NL}}\right)^{\alpha_{k}^{\textrm{NL}}/\alpha_{j}^{\textrm{NL}}}}^{\infty}\frac{\lambda_{j}^{\textrm{NL}}\left(y\right)}{1+y^{\alpha_{j}^{\textrm{NL}}}/s}\textrm{d}y\biggr],\label{eq:Proof_LT_NL_max_ave}
\end{align}
where $\widehat{\Phi_{j}^{\textrm{NL}}}=\overline{\Phi_{j}^{\textrm{NL}}}\setminus\left(0,\mathcal{Y}_{k}^{\textrm{NL}}\right]$
and in $\left(a\right)$ the lower limit of integral is $\left(\mathcal{Y}_{k}^{\textrm{NL}}\right)^{\alpha_{k}^{\textrm{NL}}/\alpha_{j}^{\textrm{NL}}}$
which guarantees that $\mathcal{P}_{k}^{\textrm{NL}}>\mathcal{P}_{j}^{\textrm{NL}},\forall j\in\mathcal{K}\setminus k$
in event E is true. Similarly, $\mathcal{L}_{I_{j}^{\textrm{L}}}^{\textrm{MARP1}}\left(s\right)$
is calculated by
\begin{align}
 & \mathcal{L}_{I_{j}^{\textrm{L}}}^{\textrm{MARP1}}\left(s\right)=\mathbb{E}_{I_{j}^{\textrm{L}}}\left[e^{-sI_{j}^{\textrm{L}}}\right]\overset{\left(a\right)}{=}\exp\biggl[-\int_{y=\left(\mathcal{Y}_{k}^{\textrm{NL}}\right)^{\alpha_{k}^{\textrm{NL}}/\alpha_{j}^{\textrm{L}}}}^{\infty}\frac{\lambda_{j}^{\textrm{L}}\left(y\right)}{1+y^{\alpha_{j}^{\textrm{L}}}/s}\textrm{d}y\biggr],\label{eq:Proof_LT_L_max_ave}
\end{align}
where in $\left(a\right)$ the lower limit of integral is $\left(\mathcal{Y}_{k}^{\textrm{NL}}\right)^{\alpha_{k}^{\textrm{NL}}/\alpha_{j}^{\textrm{L}}}$
which guarantees that $\left\{ \mathcal{P}_{k}^{\textrm{NL}}>\mathcal{P}_{j}^{\textrm{L}}\right\} ,\forall j\in\mathcal{K}$
in event E is true.

Finally, note that the value of $p_{k}^{\textrm{NL}}\left(\left\{ \lambda_{k}^{\textrm{NL}}\right\} ,\left\{ T_{k}\right\} ,\left\{ B_{k}^{\textrm{NL}}\right\} \right)$
in Eq. (\ref{eq:proof_pcNL}) should be calculated by taking the expectation
with respect to $\mathcal{Y}_{k}^{\textrm{NL}}$ in terms of its PDF,
which is given by 
\begin{equation}
f_{\mathcal{Y}_{k}^{\textrm{NL}}}\left(\varepsilon\right)=\lambda_{k}^{\textrm{NL}}\left(\varepsilon\right)\exp\left[-\Lambda_{k}^{\textrm{NL}}\left(\left[0,\varepsilon\right]\right)\right]\label{eq:PDF_y_NL}
\end{equation}
as in \cite{Yang17Density}. By substituting Eq. (\ref{eq:proof_PNL g PL}),
(\ref{eq:proof_SINR}), (\ref{eq:Proof_LT_NL_max_ave}), (\ref{eq:Proof_LT_L_max_ave}),
and (\ref{eq:PDF_y_NL}) into Eq. (\ref{eq:proof_pcNL}), we can derive
the conditional probability $p_{k}^{\textrm{NL}}\left(\left\{ \lambda_{k}^{\textrm{NL}}\right\} ,\left\{ T_{k}\right\} ,\left\{ B_{k}^{\textrm{NL}}\right\} \right)$.
Given that the typical MU is connected to a LoS BS, the conditional
coverage probability $p_{k}^{\textrm{L}}\left(\left\{ \lambda_{k}^{\textrm{L}}\right\} ,\left\{ T_{k}\right\} ,\left\{ B_{k}^{\textrm{L}}\right\} \right)$
can be derived using the similar way as the above. Thus the proof
is completed.

\section*{\lyxadded{Bean}{Wed Dec  6 12:40:26 2017}{Appendix C: Proof of Corollary
5}}

\lyxadded{Bean}{Wed Dec  6 12:58:35 2017}{By comparing $p_{\textrm{cov}}^{\textrm{MIRP}}\left(\left\{ \lambda_{k}\right\} ,\left\{ T_{k}\right\} ,\left\{ B_{k}^{\textrm{U}}\right\} \right)$
and $p_{\textrm{cov}}^{\textrm{MARP}}\left(\left\{ \lambda_{k}\right\} ,\left\{ T_{k}\right\} ,\left\{ B_{k}^{\textrm{U}}\right\} \right)$
in Theorem 1 and 2, it is noticed that the difference between them
lies in the Laplace transform and the term $e^{-\stackrel[j=1]{K}{\sum}\left[\Lambda_{j}^{\textrm{L}}\left(\left[0,r^{\alpha_{k}^{\textrm{L}}/\alpha_{j}^{\textrm{L}}}\right]\right)+\Lambda_{j}^{\textrm{NL}}\left(\left[0,r^{\alpha_{k}^{\textrm{L}}/\alpha_{j}^{\textrm{NL}}}\right]\right)\right]}$.
Thus, we prove this corollary by taking the coverage probability for
a typical MU which is served LoS BSs for an example.}

\lyxadded{Bean}{Wed Dec  6 13:09:51 2017}{
\begin{align}
 & p_{\textrm{cov}}^{\textrm{MIRP}}\left(\left\{ \lambda_{k}\right\} ,\left\{ T_{k}\right\} ,\left\{ B_{k}^{\textrm{U}}\right\} \right)>p_{\textrm{cov}}^{\textrm{MARP}}\left(\left\{ \lambda_{k}\right\} ,\left\{ T_{k}\right\} ,\left\{ B_{k}^{\textrm{U}}\right\} \right)\nonumber \\
 & \iff e^{-\stackrel[j=1]{K}{\sum}\left[\Lambda_{j}^{\textrm{L}}\left(\left[0,r^{\alpha_{k}^{\textrm{L}}/\alpha_{j}^{\textrm{L}}}\right]\right)+\Lambda_{j}^{\textrm{NL}}\left(\left[0,r^{\alpha_{k}^{\textrm{L}}/\alpha_{j}^{\textrm{NL}}}\right]\right)\right]}\stackrel[j=1]{K}{\prod}\left[\mathcal{L}_{I_{j}^{\textrm{NL}}}^{\textrm{MARP2}}\left(T_{k}r^{\alpha_{k}^{\textrm{L}}}\right)\mathcal{L}_{I_{j}^{\textrm{L}}}^{\textrm{MARP2}}\left(T_{k}r^{\alpha_{k}^{\textrm{L}}}\right)\right]/\nonumber \\
 & \quad\quad\stackrel[j=1]{K}{\prod}\left[\mathcal{L}_{I_{j}^{\textrm{NL}}}^{\textrm{MIRP}}\left(T_{k}r^{\alpha_{k}^{\textrm{L}}}\right)\mathcal{L}_{I_{j}^{\textrm{L}}}^{\textrm{MIRP}}\left(T_{k}r^{\alpha_{k}^{\textrm{L}}}\right)\right]<1\nonumber \\
 & \iff e^{-\stackrel[j=1]{K}{\sum}\left[\Lambda_{j}^{\textrm{L}}\left(\left[0,r^{\alpha_{k}^{\textrm{L}}/\alpha_{j}^{\textrm{L}}}\right]\right)+\Lambda_{j}^{\textrm{NL}}\left(\left[0,r^{\alpha_{k}^{\textrm{L}}/\alpha_{j}^{\textrm{NL}}}\right]\right)-\int_{0}^{r^{\alpha_{k}^{\textrm{L}}/\alpha_{j}^{\textrm{L}}}}\frac{\lambda_{j}^{\textrm{L}}\left(y\right)}{1+y^{\alpha_{j}^{\textrm{L}}}/T_{k}r^{\alpha_{k}^{\textrm{L}}}}\textrm{d}y-\int_{0}^{r^{\alpha_{k}^{\textrm{L}}/\alpha_{j}^{\textrm{NL}}}}\frac{\lambda_{j}^{\textrm{NL}}\left(y\right)}{1+y^{\alpha_{j}^{\textrm{NL}}}/T_{k}r^{\alpha_{k}^{\textrm{L}}}}\textrm{d}y\right]}\text{<1}\nonumber \\
 & \impliedby\Lambda_{j}^{\textrm{L}}\left(\left[0,r^{\alpha_{k}^{\textrm{L}}/\alpha_{j}^{\textrm{L}}}\right]\right)>\int_{0}^{r^{\alpha_{k}^{\textrm{L}}/\alpha_{j}^{\textrm{L}}}}\frac{\lambda_{j}^{\textrm{L}}\left(y\right)}{1+y^{\alpha_{j}^{\textrm{L}}}/T_{k}r^{\alpha_{k}^{\textrm{L}}}}\textrm{d}y,\nonumber \\
 & \quad\quad\Lambda_{j}^{\textrm{NL}}\left(\left[0,r^{\alpha_{k}^{\textrm{L}}/\alpha_{j}^{\textrm{NL}}}\right]\right)>\int_{0}^{r^{\alpha_{k}^{\textrm{L}}/\alpha_{j}^{\textrm{NL}}}}\frac{\lambda_{j}^{\textrm{NL}}\left(y\right)}{1+y^{\alpha_{j}^{\textrm{NL}}}/T_{k}r^{\alpha_{k}^{\textrm{L}}}}\textrm{d}y\nonumber \\
 & \iff\lambda_{j}^{\textrm{L}}\left(y\right)>\frac{\lambda_{j}^{\textrm{L}}\left(y\right)}{1+y^{\alpha_{j}^{\textrm{L}}}/T_{k}r^{\alpha_{k}^{\textrm{L}}}},\lambda_{j}^{\textrm{NL}}\left(y\right)>\frac{\lambda_{j}^{\textrm{NL}}\left(y\right)}{1+y^{\alpha_{j}^{\textrm{NL}}}/T_{k}r^{\alpha_{k}^{\textrm{L}}}}.
\end{align}
}

\lyxadded{Bean}{Wed Dec  6 13:10:01 2017}{The proof is completed.}

\bibliographystyle{IEEEtran}
\bibliography{BY}

\end{document}